\documentclass{IEEEtran}
\usepackage{cite}
\usepackage{amsmath,amssymb,amsfonts}
\usepackage{algorithmic}
\usepackage{graphicx}
\usepackage{textcomp}
\usepackage{amsthm}

\def\BibTeX{{\rm B\kern-.05em{\sc i\kern-.025em b}\kern-.08em
    T\kern-.1667em\lower.7ex\hbox{E}\kern-.125emX}}
\pdfoutput=1
\newtheorem{definition}{Definition}
\newtheorem{theorem}{Theorem}

\newtheorem{proposition}{Proposition}
\newtheorem{example}{Example}
\newtheorem{remark}{Remark}

\newtheorem{lemma}{Lemma}
\allowdisplaybreaks

\usepackage{balance}
\usepackage{stfloats}

\begin{document}
\title{Fundamental Limits of Bistatic Integrated Sensing and Communications over Memoryless \\Relay Channels}
\author{Yao Liu, Min Li, \IEEEmembership{Member, IEEE}, Lawrence Ong, \IEEEmembership{Senior Member, IEEE}, and Aylin Yener, \IEEEmembership{Fellow, IEEE}
\thanks{Received 5 July 2025; revised 18 November 2025; accepted 10 February 2026. The work of Min Li was supported in part by National Natural Science Foundation of China under Grant 62271440. An earlier version of this paper was presented in part at the 2024 IEEE International Symposium on Information Theory [DOI: 10.1109/ISIT57864.2024.10619459.]. (\textit{Corresponding author: Min Li}.)}
\thanks{Yao Liu is with the School of Communication Engineering, Hangzhou Dianzi University, Hangzhou 310018, China (e-mail: yao.liu@hdu.edu.cn).}
\thanks{Min Li is with the College of Information Science and Electronic Engineering and the Zhejiang Provincial Key Laboratory of Multi-Modal Communication Networks and Intelligent Information Processing, Zhejiang University, Hangzhou 310027, China (e-mail: min.li@zju.edu.cn).}
\thanks{Lawrence Ong is with the School of Engineering, University of Newcastle, Callaghan, NSW 2308, Australia (e-mail: lawrence.ong@newcastle.edu.au). }
\thanks{Aylin Yener is with the Department of Electrical and Computer Engineering, The Ohio State University, Columbus, OH 43210 USA (e-mail: yener@ece.osu.edu).}
}

\maketitle

\begin{abstract}
The problem of bistatic integrated sensing and communications over memoryless relay channels is considered, where destination concurrently decodes the message sent by the source and estimates unknown parameters from received signals with the help of a relay. A state-dependent discrete memoryless relay channel is considered to model this setup, and the fundamental limits of the communication-sensing performance tradeoff are characterized by the capacity-distortion function. An upper bound on the capacity-distortion function is derived, extending the cut-set bound results to address the sensing operation at the destination. A hybrid-partial-decode-and-compress-forward coding scheme is also proposed to facilitate source-relay cooperation for both message transmission and sensing, establishing a lower bound on the capacity-distortion function. It is found that the hybrid-partial-decode-and-compress-forward scheme achieves optimal sensing performance when the communication task is ignored. Furthermore, the upper and lower bounds are shown to coincide for three specific classes of relay channels. Numerical examples are provided to illustrate the communication-sensing tradeoff and demonstrate the benefits of integrated design.
\end{abstract}


\section{Introduction}
\par~Integrated sensing and communications (ISAC) is recognized as a promising direction for future wireless networks, enabling a number of new applications~\cite{liu2022survey,liu2022integrated}. The sharing of hardware and spectrum to perform both sensing and communication can result in a performance tradeoff between these two functionalities, due to the inherent differences between sensing-centric and communication-centric signals. Characterizing this tradeoff is thus important for gaining insights for ISAC deployment in next generation wireless networks.

\par~Previous information-theoretic investigations on ISAC ~\cite{kobayashi2018joint,liu2022information,kobayashi2019joint,ahmadipour2023information,liu2023Globecom,liu2022generalized,ahmadipour2022information,liu2024ICC,xiong2023fundamental,gunlu2023secure,welling2024transmitter,nikbakht2024integrated} have predominantly focused on monostatic scenarios, where the transmitter emits {dual-functional signals} and performs sensing estimation based on the reflected echo signals. By viewing the echo signals as output feedback and the estimated parameters as state variables, fundamental limits of monostatic point-to-point ISAC channels have been explored~\cite{kobayashi2018joint,liu2022information}, where the optimal capacity-distortion tradeoff has been derived. Such modeling methods have been further extended to multiple-access ISAC channels~\cite{kobayashi2019joint,ahmadipour2023information,liu2023Globecom,liu2022generalized}, broadcast channels~\cite{ahmadipour2022information}, interference channels~\cite{liu2024ICC}, and multi-antenna point-to-point Gaussian channels~\cite{xiong2023fundamental}, where the inner and outer bounds, or the optimal tradeoffs have been established for these respective models. In addition to the above works, security issues for monostatic ISAC have been considered~\cite{gunlu2023secure,welling2024transmitter}, while 
the problem of monostatic ISAC in the finite blocklength regime has also been investigated~\cite{nikbakht2024integrated}.

\par~Unlike monostatic sensing, bistatic sensing employs distinct nodes for the signal transmitter and the sensing estimator. In this setup, there are usually no echo signals at the transmitter. The communication receiver and the sensing estimator may be either collocated or physically separated. The studies on fundamental performance tradeoffs in such scenarios are few.
Zhang~\textit{et~al}.~\cite{zhang2011joint} were the first to consider the bistatic point-to-point and multiple-access ISAC scenarios with collocated communication receiver and sensing estimator, where one or two transmitters send dual-functional signals to enable the receiver to concurrently decode the message and perform estimation for channel states. It was shown that for these scenarios, the optimal strategy is letting the receiver to first decode the transmitters' codewords and then perform sensing through a symbol-by-symbol estimator based on the decoded codewords and received signals. Salimi~\textit{et~al}.~\cite{salimi2017capacity} further studied a two-hop line channel, considering both message decoding and state estimation at the destination, in which the optimal strategy was found to be letting the relay to first decode the codeword sent by the transmitter and then transmit the compressed version of both the decoded codeword and received signals to the receiver. 

\par~Different from Zhang~\textit{et~al}.~\cite{zhang2011joint} and Salimi~\textit{et~al}.~\cite{salimi2017capacity}, the recent studies~\cite{jiao2024rate,ahmadipour2023strong} explored the tradeoff for bistatic ISAC scenarios with physically separated communication receiver and sensing estimator. Jiao~\textit{et~al}.~\cite{jiao2024rate}
introduced a broadcast ISAC channel to model such scenarios and established the inner bound on capacity-distortion tradeoff based on the decoding-and-estimation strategies by enabling the sensing estimator to decode partial messages sent by the transmitter. Instead of considering the asymptotic case where the error probability approaches zero and the distortion constraints must be strictly satisfied, Ahmadipour~\textit{et~al}.~\cite{ahmadipour2023strong} focused on the case where the sum of the decoding error probability and the excess distortion probability is less than one and investigated the optimal rate-distortion tradeoff with the assumption that the sensing node has access to the transmitted codewords.

\par~There have also been a series of studies~\cite{sutivong2005channel,kim2008state,choudhuri2013causal,tian2015gaussian,ramachandran2019joint} focusing on the problem of joint communication and state amplification, where the transmitter has either perfect channel state information or noisy channel state information and sends dual-functional signals to assist both message decoding and parameter estimation at the receiver. These studies can be viewed as special cases of bistatic ISAC, and the results provide important insights into the fundamental limits of bistatic ISAC systems.

\par
\begin{figure}[!t]
	\centering
	\includegraphics[width=.9\linewidth]{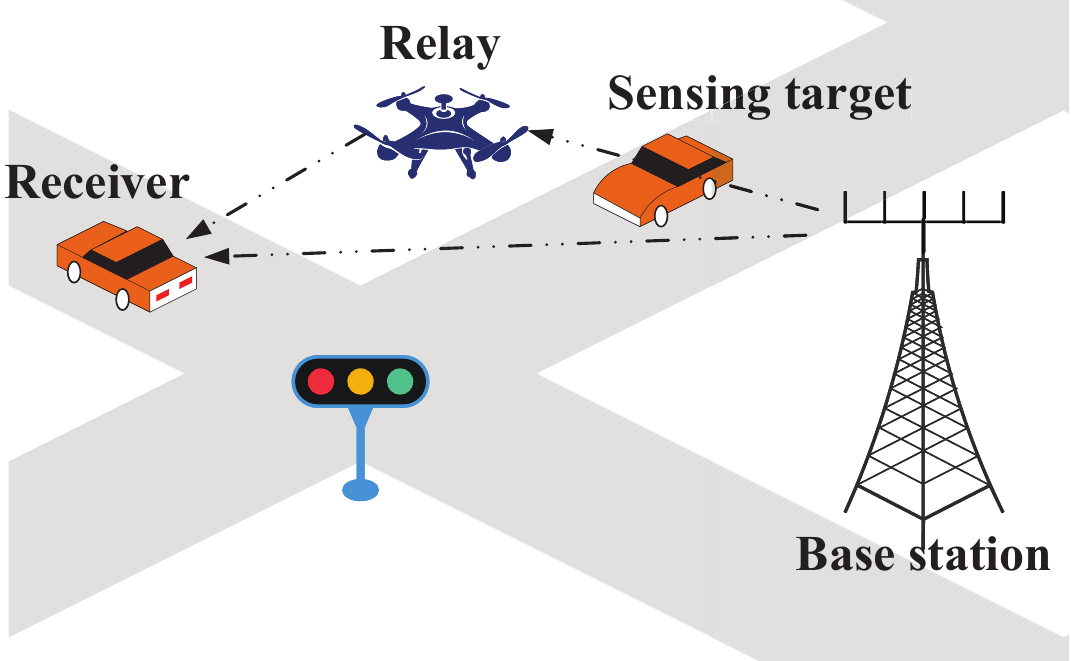}
	\caption{A motivating scenario of bistatic ISAC over relay channels.}
	\label{fig:motivating}
\end{figure}

~While the aforementioned studies\cite{zhang2011joint,salimi2017capacity,jiao2024rate,ahmadipour2023strong,sutivong2005channel,kim2008state,choudhuri2013causal,tian2015gaussian,ramachandran2019joint} have significantly advanced the understanding of bistatic ISAC in various channel settings, they primarily focus on scenarios without interactions among different nodes except for two-hop line channels~\cite{salimi2017capacity}. Future ISAC networks will involve extensive interactions among numerous nodes, including but not limited to two-hop and multi-hop scenarios. Investigating the impact of such node interactions on the performance of bistatic ISAC systems is needed for the design and optimization of ISAC systems. A canonical model is the relay-aided bistatic ISAC systems shown in Fig.~\ref{fig:motivating} in a vehicular communication network. Here, relay nodes not only extend coverage and improve transmission rates for communication, but also extract environmental information to facilitate sensing. This dual functionality of relay nodes introduces unique challenges and opportunities for the synergistic interplay between communication and sensing.

\subsection{Contributions}
\par~In this paper, we introduce a state-dependent discrete-memoryless relay channel model for a three-node bistatic ISAC system and aim to establish its capacity-distortion tradeoff. The summary of contributions is as follows:
\begin{itemize}
	\item An upper bound on the capacity-distortion function is derived, extending the cutset bound for relay channels to address the state parameter estimation at the destination. In particular, we introduce an auxiliary random variable to account for the impact of the relay on the sensing at the destination. The construction of such an auxiliary random variable is 
	inspired by joint communication and state amplification with causal channel state information at the transmitter~\cite{choudhuri2013causal}. The key difference is that in our model, the relay and the transmitter are physically separated. Therefore, the construction of the auxiliary random variable needs to consider not only the cooperation between the transmitter and the relay but also its impact on the transmission between the relay and the receiver. Our constructed auxiliary random variable can be interpreted as virtual sensing information obtained by the relay with full cooperation of the source. Such virtual sensing information is beneficial for the sensing. However, it comes at the cost of a reduction in communication rate, as the resources used to convey this additional sensing information may be diverted from the primary communication task. As a result, in addition to the sensing distortion constraint, it brings additional cost terms in traditional cut-set bound on communication rate.
	\item A lower bound on the capacity-distortion function is proposed, building upon the hybrid-partial-decode-and-compress-forward coding scheme. The introduced coding scheme involves the combined use of rate-splitting, compression without binning, block Markov encoding, and backward decoding, which are designed to facilitate source-relay cooperation for both message transmission and parameter estimation. The hybrid relaying scheme was proposed by Cover and El~Gamal~\cite{cover1979capacity} by leveraging the block Markov encoding and successive decoding. Chong~\emph{et~al.}~\cite{chong2006generalized} later proposed a new backward decoding strategy for this scheme, where in each block, the destination first performs joint decoding for the indices of common message (decoded by both the relay and the destination) and the compressed information sent by the relay, and then decodes the index of private message (decoded by the destination only). In this paper, we enhance the decoding strategy by enabling the destination to simultaneously decode the indices of common message, private message, and the compressed information sent by the relay. We show that doing so can strictly enlarge the achievable rate-distortion region. 
	\item Based on the derived upper and lower bounds, we pinpoint the sensing optimal point, i.e., the minimum distortion for capacity-distortion tradeoff region. Specifically, block Markov encoding, in which the source sends the pilot signal and the relay communicates a description of the received signals in the previous block by incorporating the side information at both the relay and the destination, achieves the optimal sensing performance. We further demonstrate two simpler strategies for the relay, which can also achieve the optimal sensing performance for two classes of relay channels, respectively.
	\item We show that the established upper and lower bounds coincide for bistatic ISAC over specific classes of relay channels, thereby characterizing the optimal capacity-distortion function. In particular, we provide three classes of relay channels, where partial-decode-forward strategy, compress-forward strategy, and hybrid-partial-decode-and-compress-forward strategy achieve the optimal capacity-distortion functions, respectively.
\end{itemize}

\subsection{Organization and Notation}
\par
\begin{figure*}[!t]
	\centering
	\includegraphics[width=.6\linewidth]{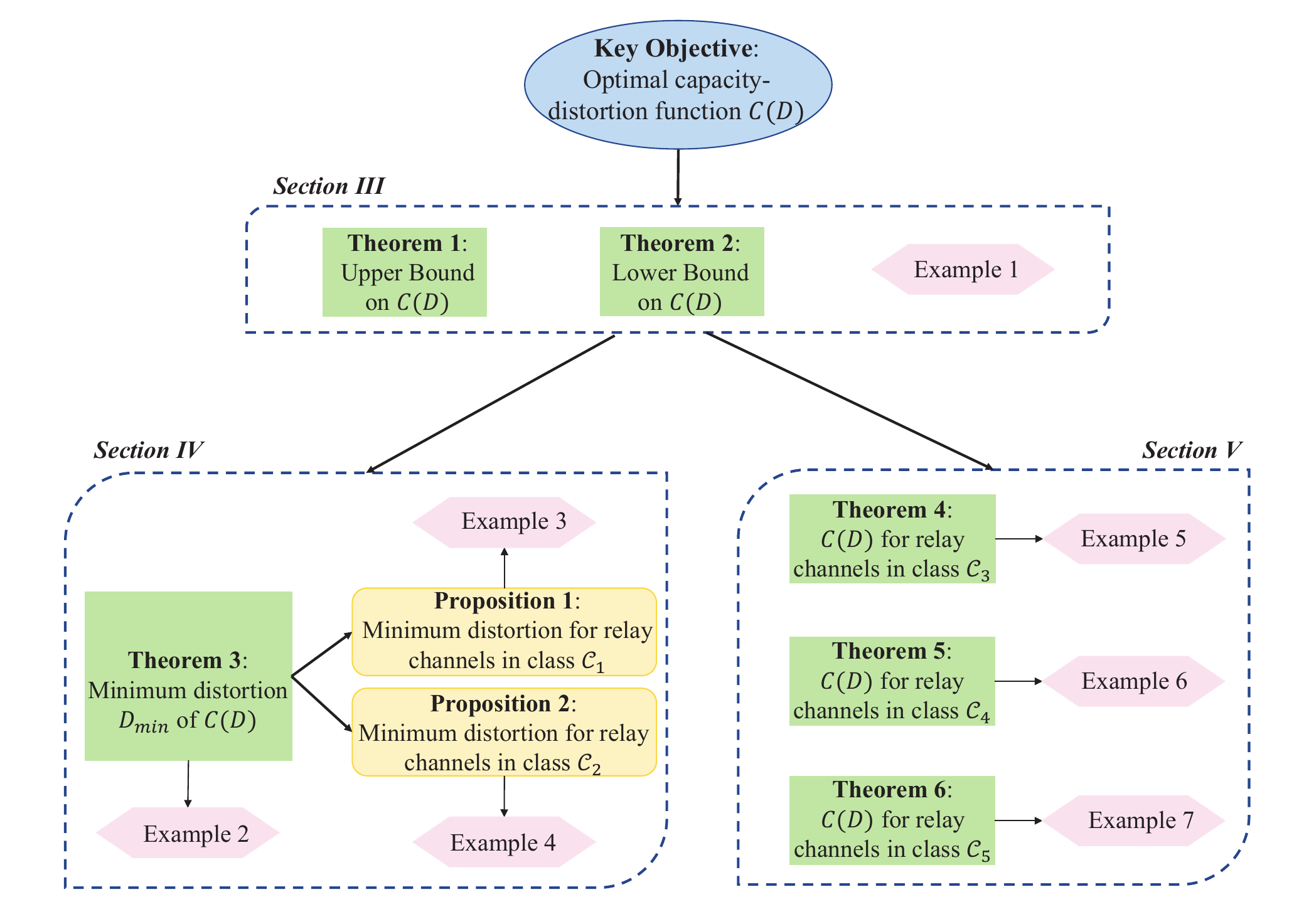}
	\caption{The logical connections between theorems, propositions, and numerical examples in this paper.}
	\label{fig:theoremRelationship}
\end{figure*}

~The remainder of this paper is organized as follows. Section~\ref{section:systemModel} describes the model for bistatic ISAC over state-dependent discrete memoryless relay channels considered in this work. Section~\ref{sec:outerAndInner} presents the main results of upper and lower bounds on the capacity-distortion function, as well as the comparison with related works. Section~\ref{sec:pureSensing} provides the optimal sensing performance, i.e., minimum distortion of capacity-distortion tradeoff region. Section~\ref{sec:optimalCD} presents the optimal capacity-distortion function for three specific classes of relay channels. Section~\ref{sec:conclusion} concludes the paper. Fig.~\ref{fig:theoremRelationship} demonstrates the logical connections among theorems, propositions, and numerical examples.

\par~\emph{Notation}: Throughout the paper, we use calligraphic letters, uppercase letters, and lowercase letters to denote sets, random variables, and the realizations, respectively, e.g., $\mathcal{X},X,x$. The probability distributions are denoted by $P$ with the subscript indicating the corresponding random variables, e.g., $P_X(x)$ and $P_{Y|X}(y|x)$ are the probability of $X=x$ and conditional probability of $Y=y$ given $X=x$. We use $x^i$ to denote the vector $[x_1,x_2,\cdots,x_i]$, $[1:L]$ to denote the set $\{1,2,\cdots,L\}$ for integer $L$, and $\mathbb{E}(X)$ to denote the expectation of random variable $X$. Logarithms are taken with respect to base 2. For binary random variables $X$, we use $P_X\triangleq P_{X}(x=1)$ to denote the probability of $X=1$. Function $p_1*p_2$, $H_2(t)$, $H_3(t)$ are defined as $p_1*p_2 \triangleq p_1(1-p_2)+(1-p_1)p_2$, $H_2(t)\triangleq-t\log_2(t)-(1-t)\log_2(1-t)$, and $H_3(t_1,t_2)\triangleq-t_1\log_2(t_1)-t_2\log_2(t_2)-(1-t_1-t_2)\log_2(1-t_1-t_2)$. $\mathcal{N}(0,\sigma^2)$ denotes a zero-mean Gaussian distribution with variance $\sigma^2$. 

\section{System Model}\label{section:systemModel}
\par
\begin{figure}[!t]
	\centering
	\includegraphics[width=.99\linewidth]{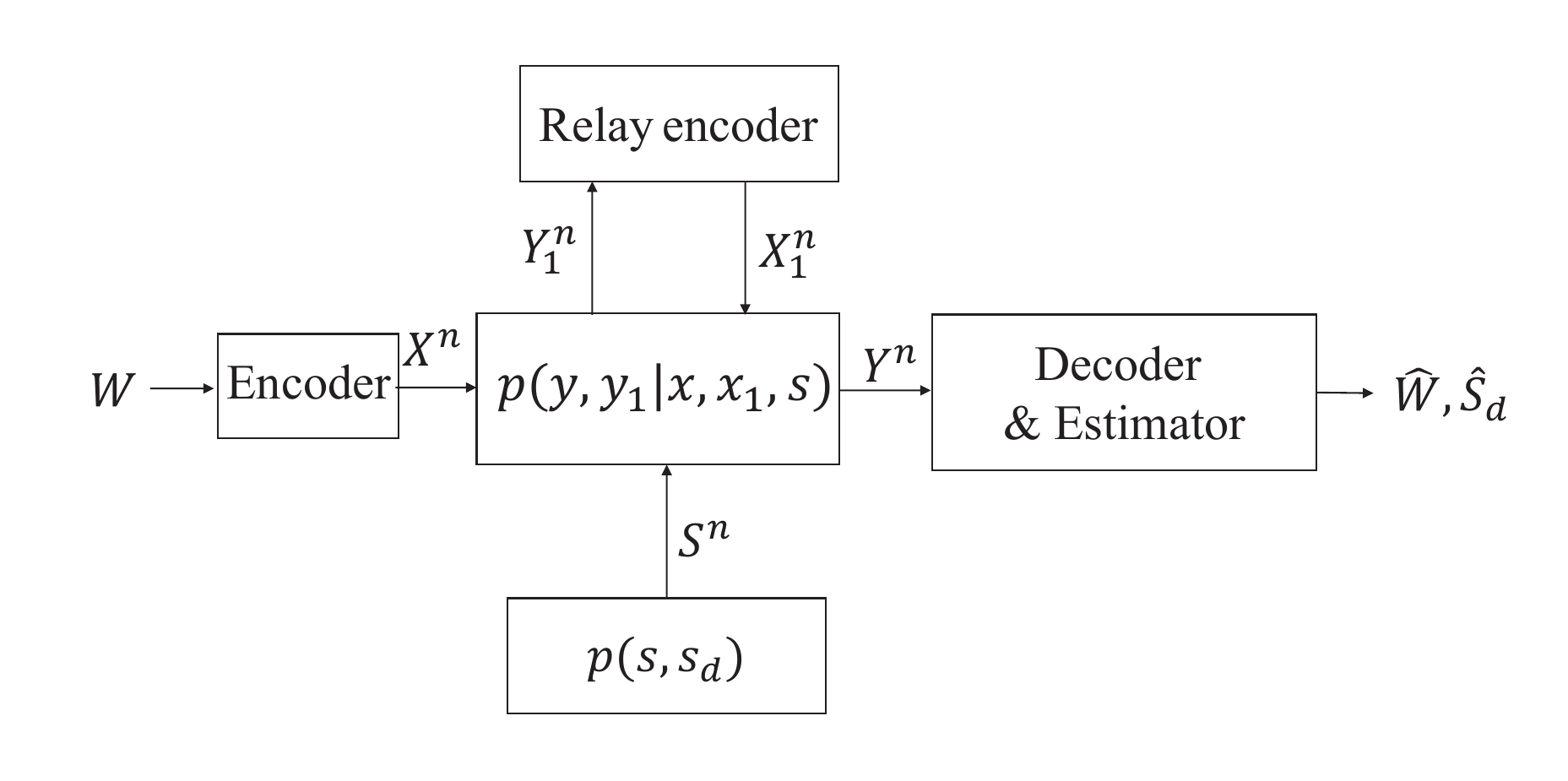}
	\caption{Bistatic ISAC over SD-DM relay channels.}
	\label{fig:channel}
\end{figure}

~Consider a bistatic ISAC system over a state-dependent discrete memoryless relay channel as shown in Fig.~\ref{fig:channel}. The source wishes to convey a message $W\in[1:2^{nR}]\triangleq\{1,2,\dotsc,2^{\lceil nR\rceil}\}$ to the destination with the help of the relay over $n$ channel uses. In addition to message decoding, the destination also aims to estimate the state parameters $S_{d}^n\triangleq [S_{d,1},\dotsc,S_{d,n}]$ that is correlated to the channel state sequence $S^n$, where $S_{d,i}$ and $S_i$ denote the estimated parameter and channel state during channel use $i\in[1:n]$. In general, the state of the channel directly (and physically) influences the channel output, and state parameters are the information that the receiver tries to capture. For example, the channel state could be the channel attenuation, and the estimated parameters may be the relative velocity of the receiver with respect to the transmitter.

\par~The bistatic ISAC relay channel is mathematically represented by 
\begin{align}
	(\mathcal{X}\times\mathcal{X}_1,\mathcal{S}\times\mathcal{S}_{d},\mathcal{Y}\times\mathcal{Y}_1,P_{YY_1|XX_1S}P_{SS_{d}}) 
\end{align}
with input alphabets $\mathcal{X}\times\mathcal{X}_1$, state alphabet $\mathcal{S}$, sensing parameter alphabet $\mathcal{S}_{d}$, output alphabets $\mathcal{Y}\times\mathcal{Y}_1 $, and a collection of probability mass functions (pmfs) $P_{YY_1|XX_1S}P_{SS_{d}}$. The joint distribution of channel state $S$ and sensing parameter $S_{d}$ is given by $P_{SS_{d}}$, which is independent and identically distributed (i.i.d.) according to 
\begin{align}
	P_{S^nS_{d}^n} = \prod_{i=1}^nP_{SS_{d}}(s_i,s_{{d},i}).
\end{align}
The channel is memoryless in the sense that 
\begin{align}
	P(y_i,y_{1,i}|x^i,x_1^i,s^i,y^{i-1},y^{i-1}_1) = P(y_i,y_{1,i}|x_i,x_{1,i},s_{i}), 
\end{align}
for $i\in[1:n]$, where $x_i,x_{1,i}$ and $y_i,y_{1,i}$ are channel inputs and outputs during channel use $i\in[1:n]$, respectively.

\begin{definition}
	A $(2^{nR},n)$ code for bistatic ISAC relay channel consists of 
	\begin{enumerate}
		\item A message set $\mathcal{W}=[1:2^{nR}]$, where message $W$ is uniformly distributed over the message set $\mathcal{W}$.
		\item An encoder at the source that assigns an input sequence $x^n=e_n(w)$ to each message $w\in\mathcal{W}$.
		\item An encoder at the relay that assigns a symbol $x_{1,i}=f_i(y_{1}^{i-1})$ to the past received sequence $y_{1}^{i-1}\in\mathcal{Y}_1^{i-1}$, $i\in[1:n]$.
		\item A decoder at the destination that produces an estimated message $\hat{w}=h_n(y^n)\in\mathcal{W}$ from the received sequence $y^n\in\mathcal{Y}^n$.
		\item A state estimator at the destination that produces an estimated parameter sequence $\hat{s}_{d}^n=g_n(y^n)\in\mathcal{S}_{d}^n$ from the received sequence $y^n\in\mathcal{Y}^n$.
	\end{enumerate}
\end{definition}

\par~The sensing performance is measured by the expected distortion of the estimated parameters, i.e., 
\begin{align}\label{equ:defDistortionBlockN}
	\Delta^{(n)}=\mathbb{E}[d(S_{d}^n,\hat{S}_{d}^n)] = \frac{1}{n}\sum_{i=1}^n \mathbb{E}[d(S_{{d},i},\hat{S}_{{d},i})], 
\end{align}
where $d:\mathcal{S}_{d}\times\hat{\mathcal{S}}_{d}\rightarrow[0,\infty)$ is a bounded distortion function. 

\begin{definition}
	A rate-distortion pair $(R,D)$ is said to be achievable if there exists a sequence of $(2^{nR},n)$ codes with diminishing message-decoding error probability, i.e., 
	\begin{align}
		\lim_{n\rightarrow\infty}\Pr(\hat{W}\neq W) = 0, 
	\end{align}
	and the sensing distortion constraint 
	\begin{align}
		\limsup_{n\rightarrow\infty}\mathbb{E}[d(S_{d},\hat{S}_{d})]\le D  
	\end{align}
	is satisfied. The minimum distortion $D_{\text{min}}$ is defined as the minimum value of $D$ such that there exists some $R$ makes rate-distortion pair $(R,D)$ achievable.
	For any given $D\ge D_{\text{min}}$, the capacity-distortion function $C(D)$ is defined as the supremum of $R$ such that the rate-distortion pair $(R,D)$ is achievable.  
\end{definition}

\par~Our goal is to characterize the capacity-distortion function $C(D)$ for bistatic ISAC over relay channels considered in Fig.~\ref{fig:channel}. Such a goal is extremely challenging since even without sensing tasks, the capacity of general relay channels remains unknown. Therefore, in the following sections of the paper, we first provide the general upper and lower bounds for $C(D)$, then identify the optimal sensing performance $D_{min}$ for $C(D)$, and finally provide $C(D)$ for three specific classes of relay channels.

\section{Upper and Lower Bounds on Capacity-Distortion Function $C(D)$}\label{sec:outerAndInner}
\subsection{Upper Bound on $C(D)$}
\par~To establish an upper bound on $C(D)$, we extend the cutset bound results for the relay channels~\cite{el2011network} to address the parameter estimation at the destination. This is achieved by introducing an auxiliary random variable $T$ to characterize the impact of relay on the performance tradeoff between sensing and communication at the destination. In particular, inspired by the construction of auxiliary random variable for joint communication and state amplification with causal channel state information at the transmitter~\cite{choudhuri2013causal}, we introduce the auxiliary random variable $T$ to characterize the useful information at the relay for sensing, assuming full cooperation between the source and the relay. Such information is sent to the destination for sensing, which also incurs a loss in communication performance. Accordingly, we derive the transmission constraint for sending the useful information to the destination, the loss in communication rate, and also the sensing distortion constraint. Combined with the cutset bound results on communication rate, we thus establish the upper bound of $C(D)$.

\begin{theorem}\label{theorem:outer}
	The capacity-distortion function $C(D)$ is upper bounded as
	\begin{align}\label{equ:rateUpperBoundCD}
		C(D)\le &\max_{P_{XX_1}P_{T|XX_1Y_1}\in\mathcal{P}_D}\min\bigg\{I(X;YY_1|X_1),\notag\\
		&\qquad\qquad\quad
		I(XX_1;Y)-I(T;Y_1|XX_1Y)\bigg\},
	\end{align}
	where 
	\begin{align}
		\mathcal{P}_D = \bigg\{&P_{XX_1}P_{T|XX_1Y_1}:\mathbb{E}[d(S_{d},\hat{S}_{d})]\le D,\notag\\
		&\qquad\quad~{\text{and}}~ I(X_1;Y|X)\ge I(T;Y_1|XX_1Y)\bigg\},
	\end{align}
	$\hat{S}_d$ is a deterministic function of $X,X_1,Y,T$ given as
	\begin{align}\label{equ:stateEstimator}
		\hat{S}_{d}(x,&x_1,y,t) =\notag\\
		& \operatorname*{argmin}_{s'_{d}\in\mathcal{S}_{d}}\sum_{s_{d}\in\mathcal{S}_{d}}P_{S_{d}|XX_1YT}(s_{d}|x,x_1,y,t)d(s_{d},s'_{d}),
	\end{align}
	and the joint distribution of variables $XX_1SS_{d}YY_1T\hat{S}_{d}$ is 
	\begin{align}
		P_{XX_1}P_{SS_{d}}P_{YY_1|XX_1S}P_{T|XX_1Y_1}P_{\hat{S}_{d}|XX_1YT},
	\end{align}
	where $P_{SS_{d}}P_{YY_1|XX_1S}$ is fixed by the channel, $P_{\hat{S}_{d}|XX_1YT}$ is fixed by the chosen estimator~\eqref{equ:stateEstimator}, and $P_{XX_1}P_{T|XX_1Y_1}$ is to be optimized. It suffices to consider $T$ whose alphabet $\mathcal{T}$ has cardinality $|\mathcal{T}|\le |\mathcal{X}||\mathcal{X}_1|+1$.
	\begin{IEEEproof}
		See Appendix~\ref{appendix:proofOuter}.
	\end{IEEEproof}
\end{theorem}

\begin{remark}
	As stated above, the auxiliary random variable $T$ can be interpreted as the useful information that the relay extracts for sensing at the receiver. $T$ is generated according to the distribution $P_{T|XX_1Y_1}$, which indicates that such an upper bound is under the assumption of full cooperation between the source and the relay. Such information $T$ is sent to the receiver for sensing, as evidenced by the estimator~\eqref{equ:stateEstimator}. The corresponding transmission constraint is $I(X_1;Y|X)\ge I(T;Y_1|XX_1Y)$, and the cost in communication rate is represented by term $I(T;Y_1|XX_1Y)$ in the multiple-access cut in rate constraints~\eqref{equ:rateUpperBoundCD}.
\end{remark}

\begin{remark}
	When there is no sensing task, the results in Theorem~\ref{theorem:outer} specializes to the cutset bound for the relay channels~\cite{el2011network}.
\end{remark}

\subsection{Lower Bound on $C(D)$}
\par~To establish a lower bound on $C(D)$, we propose a hybrid-partial-decode-and-compress-forward coding scheme, which is built on a coding scheme with block Markov encoding and backward decoding. In each block, the source splits the messages into a common part and a private part. The relay decodes the common part of messages sent from the source and then transmits both the decoded common message and the compressed information about received signals to the destination. The key difference between our achievable scheme and the scheme proposed by Chong, Motani, and Garg in~\cite{chong2006generalized} (We refer to this scheme as the Chong-Motani-Garg scheme) is that in backward decoding procedures, our scheme lets the destination simultaneously decode the indices of common message, private message, and the compressed information, while in Chong-Motani-Garg scheme~\cite{chong2006generalized}, the destination first performs jointly decoding for indices of common message and the compressed information, and then decodes the index of private message. The benefits of simultaneously decoding these three indices are discussed in Remark~\ref{remark:discussSimultaneouslyDecodingAdvantage} below. After backward decoding, the destination performs state estimation with received signal $Y$, decoded signals $X,X_1$, and compressed information 
sent from the relay. 

\par
\begin{figure}[!t]
	\centering
	\includegraphics[width=.99\linewidth]{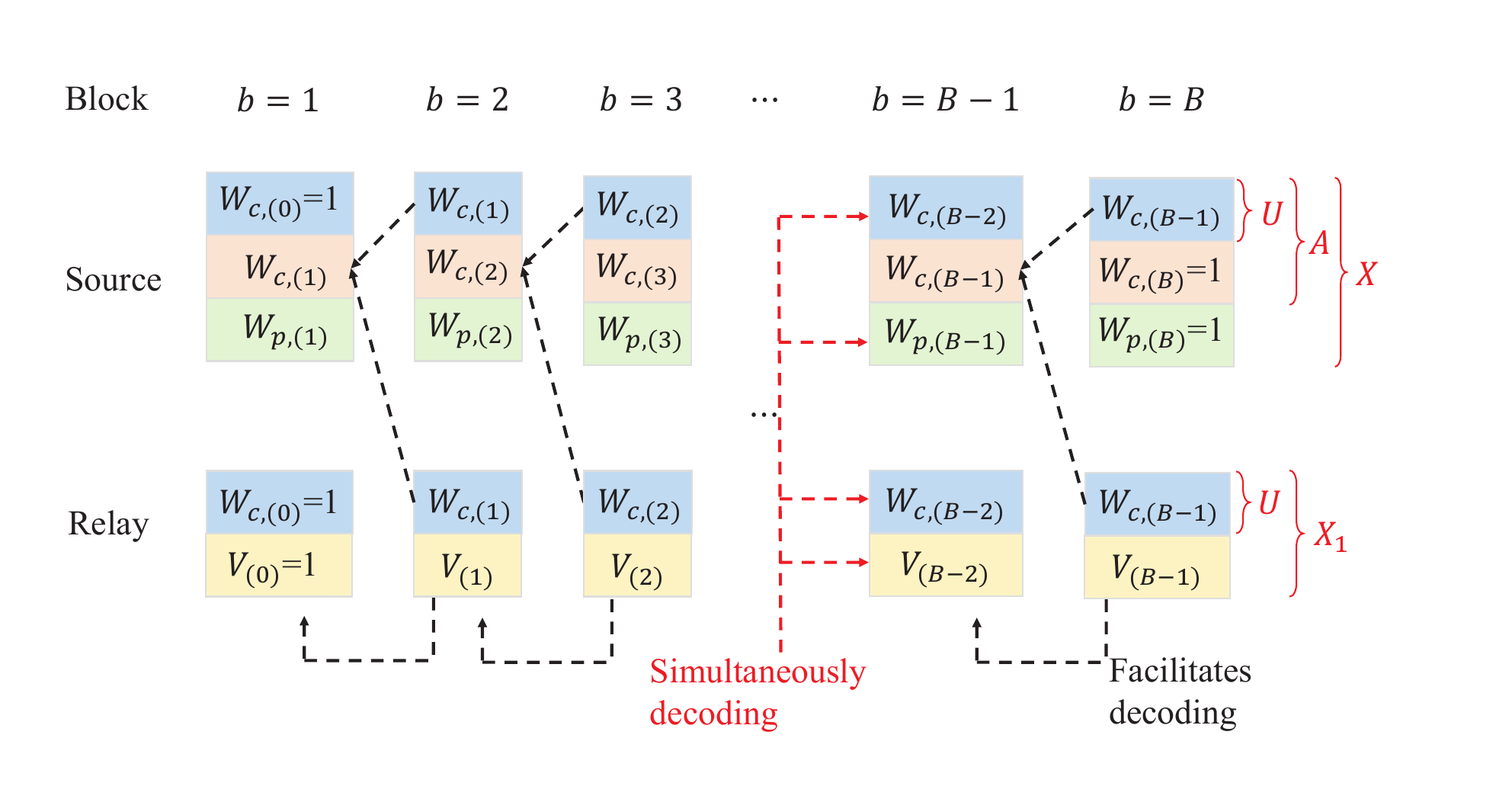}
	\caption{An illustration of our achievable coding scheme.}
	\label{fig:achievable_scheme}
\end{figure}

~An illustration of our achievable scheme is shown in Fig.~\ref{fig:achievable_scheme}, 
which involves auxiliary random variables $U,A,V$. Specifically, $U$ denotes the cooperative signal corresponding to the common message sent by the source in the previous block. $A$ denotes the combination of common message $U$ and the fresh common message sent by the source in the current block. $V$ denotes the compressed information forwarded by the relay. $X$ and $X_1$ are the actual transmitted signals at the source and the relay, respectively. Here, $X$ includes $A$ together with the fresh private message sent by the source in the current block, while $X_1$ combines the common message $U$ and the compressed information $V$. The corresponding lower bound is summarized in the theorem below.

\begin{theorem}\label{theorem:inner}
	The capacity-distortion function $C(D)$ is lower bounded by any nonnegative rate $R$ satisfying 
	\begin{subequations}\label{equ:ourRateCons}
		\begin{align}
			R \le& I(A;Y_1|UX_1) + I(X;VY|UAX_1), \label{equ:ourRateConsOne} \\
			R \le& I(A;Y_1|UX_1) + I(XX_1;Y|UA) - I(V;Y_1|UAXX_1Y),\label{equ:ourRateConsTwo}\\
			R \le& I(XX_1;Y) - I(V;Y_1|UAXX_1Y),  \label{equ:ourRateConsThree}
		\end{align}
	\end{subequations}
	and 
	\begin{align}\label{equ:ourConstraint}
		I(X_1;Y|UAX) \ge I(V;Y_1|UAXX_1Y), 
	\end{align}
	as well as the sensing distortion constraint
	\begin{align}\label{equ:ourDisConstraint}
		\mathbb{E}[d(S_{d},\hat{S}_{d}(X,X_1,Y,V))]\le D, 
	\end{align}
	where estimator~\eqref{equ:stateEstimator} is considered, and the joint distribution of $UAXX_1SS_{d}YY_1V\hat{S}_{d}$ is 
	\begin{align}
		P_UP_{A|U}P_{X|UA}P_{X_1|U}P_{SS_{d}}&P_{YY_1|XX_1S}\notag\\
		&
		P_{V|UAX_1Y_1}P_{\hat{S}_{d}|XX_1YV},
	\end{align}
	where $P_{SS_{d}}P_{YY_1|XX_1S}$ is fixed by the channel, $P_{\hat{S}_{d}|XX_1YV}$ is fixed by the estimator~\eqref{equ:stateEstimator}, and $P_UP_{A|U}P_{X|UA}P_{X_1|U}$ $P_{V|UAX_1Y_1}$ is to be optimized.
	\begin{IEEEproof}
		See Appendix~\ref{appendix:proofInner}.
	\end{IEEEproof}
\end{theorem}

\begin{remark}
		The lower-bound results in Theorem~\ref{theorem:inner} can be interpreted as follows: 
		\begin{itemize}
			\item For rate constraints in~\eqref{equ:ourRateConsOne}--\eqref{equ:ourRateConsThree}, the term $I(A;Y_1|UX_1)$ represents the rate constraint of common message that can be reliably decoded by the relay. The terms $I(X;VY|UAX_1)$ and $I(XX_1;Y|UA)$ correspond to the rate constraints of the private message that can be reliably decoded by the destination in two scenarios to be explained next. The first term applies when the destination has already decoded the common message and the compressed information, whereas the second term applies when the destination jointly decodes the private message and the compressed information. The term $I(XX_1;Y)$ represents the rate constraint for the entire message, where the destination simultaneously decodes the common message, private message, and compressed information. Finally, $I(V;Y_1|UAXX_1Y)$ quantifies the reduction in available
			message rate due to the transmission of the compressed information.
			\item Inequality constraint~\eqref{equ:ourConstraint} is imposed to ensure
			the successful transmission of the compressed information from the relay to the destination, where the term $I(X_1;Y|UAX)$ represents the achievable rate of the relay-destination link when the destination has already decoded the overall message.
			\item Inequality constraint~\eqref{equ:ourDisConstraint} represents the distortion constraint for sensing, where the destination preforms symbol-by-symbol estimator based the decoded transmitted signals $X,X_1$, received signal $Y$, and the compressed information $V$.
		\end{itemize}
		While the upper bound in Theorem~\ref{theorem:outer} assumes full cooperation between the source and the relay, the lower bound here is achieved under a practical hybrid-partial-decode-and-compress-forward scheme, which introduces additional mutual-information penalties as expressed in~\eqref{equ:ourRateCons}--\eqref{equ:ourDisConstraint}. Consequently, the gap between the upper and lower bounds quantifies the performance loss incurred by practical relying and decoding strategies, compared to ideal unconstrained joint processing.
\end{remark}

\begin{remark}\label{remark:discussSimultaneouslyDecoding}
	For communication over relay channels, the Chong-Motani-Garg scheme~\cite{chong2006generalized} considered two backward decoding strategy:
	\begin{itemize}
		\item In the first strategy, the destination decodes the indices of common message, compressed information, and private message in a sequential manner. 
		\item In the second strategy, the destination first decodes the indices of common message and compressed information simultaneously, and then decodes the index of private message.  
	\end{itemize}
	It is shown that with the same block Markov encoding procedures, the achievable rate for the second strategy includes that of the first strategy. However, whether this inclusion is strict remains open~\cite{chong2006generalized}. This open problem also appears for the communication over interference channels~\cite{han1981new}. In this paper, we always refer to the second backward decoding strategy as the Chong-Motani-Garg scheme~\cite{chong2006generalized}. 
\end{remark}

\begin{remark}\label{remark:discussSimultaneouslyDecodingAdvantage}
	It is also not clear whether our proposed simultaneously decoding for three indices strictly improves the achievable rate for pure message communication compared with the Chong-Motani-Garg scheme~\cite{chong2006generalized}. However, we demonstrate in Appendix~\ref{appendix:discussionSimultaneouslyDecoding} that this strategy does strictly enlarge the achievable rate-distortion region for ISAC over relay channels. The rationale is as follows. In the Chong-Motani-Garg scheme~\cite{chong2006generalized}, the destination simultaneously decodes the indices of the common message and the compressed information, while treating the private message as interference. As a result, the compressed information must be decoded in a noisier condition, which reduces its reliability and degrades the sensing performance. In contrast, our proposed scheme performs a joint typicality test over all three components (common message, private message, and compressed information), thereby fully exploiting the received superposition. This approach effectively turns the private message from interference into additional useful observations, enhancing the signal-to-noise ratio for retrieving the compressed information and consequently reducing the sensing distortion. 
	This improvement aligns with a well-known principle in multiple-access channel: joint decoding achieves the full capacity region, whereas successive decoding treats part of the signal as interference and generally attains smaller achievable region.
\end{remark}

\subsection{A Numerical Example}
\begin{example}\label{example:generalAchievable}
	Consider a relay channel with binary inputs $X,X_1\in\{0,1\}$. The channel output at the relay is $Y_1=S_1X$, and the channel output at the destination is $Y=S_2X+S_3X_1$, where the channel states $S_1,S_2,S_3$ are mutually independent binary random variables. The sensing parameter at the destination is $S_{d}=S_1$, and the Hamming distortion $d(s_{d},\hat{s}_{d})=s_{d}\oplus\hat{s}_{d}$ is considered. 
	
	For this example, we evaluate the lower bound specified in Theorem~\ref{theorem:inner} by considering a specific choice of joint distribution $P_UP_{A|U}P_{X|UA}P_{X_1|U}P_{V|UAX_1Y_1}$ with 
	\begin{align}
		&X = A \oplus \Theta = U \oplus \Sigma \oplus \Theta, \quad X_1 = U \oplus \Delta, \quad V = NS_1X, \notag
	\end{align}
	where $U$, $\Sigma$, $\Theta$, $\Delta$, and $N$ are mutually independent binary random variables.
	
	Regarding the upper bound, direct evaluation as per Theorem~\ref{theorem:outer} becomes intricate due to the presence of auxiliary random variable with a large cardinality that hinders the exhaustive search. Even for the example considered with binary input $X,X_1\in\{0,1\}$, given the joint distributions $P_{XX_1}P_{T|XX_1Y_1}$ to be optimized, there are up to $27$ variables to be optimized. Similar complexity arises in the characterization of dependence-balanced outer bounds for communication over multiple access channels with feedback~\cite{tandon2009outer}. For a more manageable numerical assessment, we opt for a relaxed upper bound as
		\begin{align}
			C(D) \le \max_{P_{XX_1}P_{\hat{S}_{d}|S_{d}}} \min\{I(X;YY_1|X_1),I(XX_1;Y)\}, 
		\end{align}
		subject to the constraints:
		\begin{align}
			&I(S_{d};\hat{S}_{d})\le \min\{I(S_{d};YY_1|XX_1),I(S_{d}X_1;Y|X)\},\\
			&\mathbb{E}[d(S_{d},\hat{S}_{d})]\le D. 
		\end{align}
		The proof of the relaxation is provided in Appendix~\ref{appendix:relaxationOFouterBound}.
 
\begin{figure}[!t]
	\centering
	\includegraphics[width=.99\linewidth]{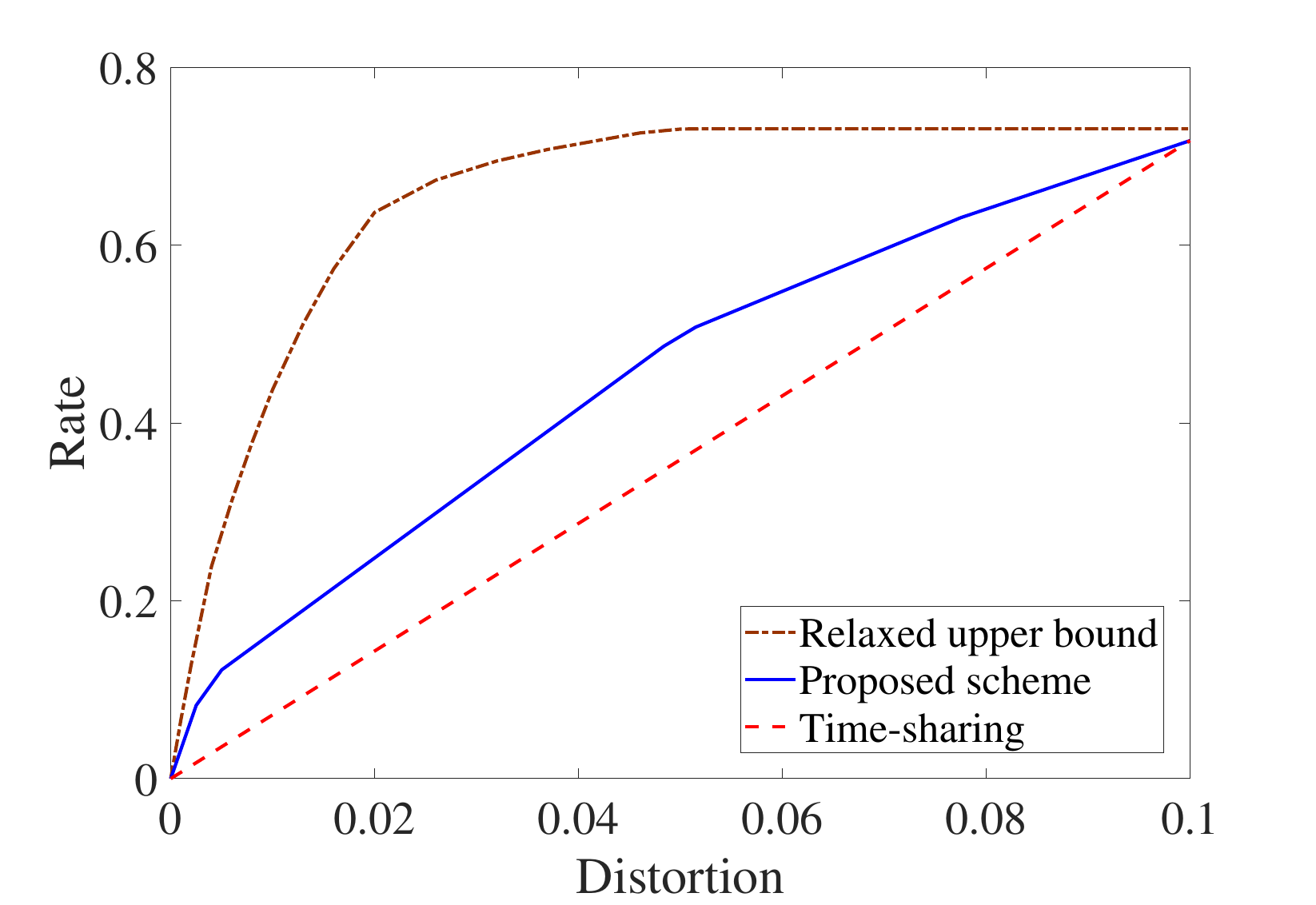}
	\caption{Rate-distortion tradeoff for the channel considered in Example~\ref{example:generalAchievable}.}
	\label{fig:ExampleOne}
\end{figure}
	
	The numerical results of the relaxed upper bound and the lower bound are illustrated in Fig.~\ref{fig:ExampleOne} for the given probabilities  $P_{S_1} =0.9$, $P_{S_2}=0.1$, and $P_{S_3}=0.9$. Notably, the proposed coding scheme achieves sensing performance with $D_{\text{min}}=0$ when optimized for sensing, and achieves communication performance of $R_{\text{max}}=0.7177$ when optimized for communication, which is close to the upper bound $R_{\text{upper,max}}=0.73$.
	In addition, our proposed scheme delivers a noticeable improvement, in comparison with the time-sharing approach, where the system alternates between $(D,R) = (0,0)$ and $(0.1,0.7177)$.

	It is also worth noting that the derived lower bound is sub-optimal except for specific performance point and channel models, which is shown in the following sections. 
\end{example}

\section{Minimum Distortion of Capacity-Distortion Tradeoff Region}\label{sec:pureSensing}
\par~The capacity-distortion function $C(D)$ delineates the Pareto frontier between the communication rate and sensing accuracy. The leftmost point of the Pareto frontier corresponding to the minimum distortion $D_{\text{min}}$, represents a fundamental performance limit in its own right: it characterizes the lowest parameter estimation error when the communication task is disregarded. In other words, $D_{\text{min}}$ is analogous to the ``sensing capacity'' of the channel with respect to the distortion metric. Investigating $D_{\text{min}}$ is therefore essential, not only to reveal the extent of sensing accuracy that must be sacrificed to achieve a given communication rate, but also to provide insights for subsequent analyses of the optimal tradeoff $C(D)$ under specific channel conditions. 

\par~To this end, based on the established upper and lower bounds, we now study the minimum distortion $D_{\text{min}}$ of capacity-distortion tradeoff region
in this section. 
Specifically, we first characterize the minimum distortion for the general relay model in the following theorem.   
\begin{theorem}\label{theorem:optimalSensing}
	The minimum distortion of capacity-distortion tradeoff region is 
	\begin{align}\label{equ:minimumDistortion}
		D_{\text{min}} = \min_{P_{XX_1}P_{V|XX_1Y_1}}\mathbb{E}[d(S_{d},\hat{S}_{d})], 
	\end{align}
	where the minimum is over all pmfs $P_{XX_1}P_{V|XX_1Y_1}$ such that 
	\begin{align}\label{equ:descriptionCons}
		I(X_1;Y|X)\ge I(V;Y_1|XX_1Y), 
	\end{align}
	$\hat{S}_{d}$ is a deterministic function of $XX_1YV$ given in~\eqref{equ:stateEstimator}, and the joint distribution of variables $XX_1SS_{d}YY_1V\hat{S}_{d}$ is 
	\begin{align}
		P_{XX_1}P_{SS_{d}}P_{YY_1|XX_1S}P_{V|XX_1Y_1}P_{\hat{S}_{d}|XX_1YV}. 
	\end{align}
	\begin{IEEEproof}
		The converse proof is tailored from that of Theorem~\ref{theorem:outer} by omitting the consideration of communication performance, setting $T=V$, and optimizing $P_{XX_1}P_{T|XX_1Y_1}=P_{XX_1}P_{V|XX_1Y_1}$ to obtain a lower bound on sensing distortion. Regarding achievability, setting $U=A=X$ in Theorem~\ref{theorem:inner}, is adequate to find that the minimum distortion $D_{\text{min}}$ can be achieved. Therefore, $D_{\text{min}}$ characterizes the minimum distortion over all achievable rate-distortion pair $(R,D)$ for bistatic ISAC over general relay channels.
	\end{IEEEproof}
\end{theorem}

\begin{remark}\label{remark:optimalSensing}
	To achieve the minimum distortion as shown in Theorem~\ref{theorem:optimalSensing}, the input distributions of $X,X_1$, i.e., $P_{XX_1}$, should be jointly optimized, and the relay should exactly know the input of the source so that it can choose the random variable $V$ according to distortion $P_{V|XX_1Y_1}$. 
	One straightforward approach to achieving the minimum distortion is for the source to transmit pilot signals such that the relay can perfectly recover the input $X$ in each block. In the next block, the relay performs the joint compression of the pilot signal $X$, its own transmitted signal $X_1$, and received signal $Y_1$ corresponding to the last block. Such compressed information $V$ is sent to the destination to facilitate the parameter estimation. While using pilots is one simple approach to attain the minimum distortion, other achievable coding schemes may exist that can simultaneously convey information and achieve the same minimum distortion for certain classes of ISAC relay channels. One such example is Example~\ref{example:optimalOrthogonalInputOrthogonalOutput} to be studied later in Section~\ref{sec:optimalCD}.
\end{remark}
	
\par~To illustrate the results in Theorem~\ref{theorem:optimalSensing}, we consider the following numerical example:
\begin{example}\label{example:optimalSensingGeneral}
	Consider a Gaussian relay channel with additive Gaussian states 
	\begin{align}
		Y_1 = X + S_1, \ Y = X + X_1 + S_1 + S_2,
	\end{align}
	where $S_1\sim N(0,\sigma^2_{s_1})$ and $S_2\sim N(0,\sigma^2_{s_2})$ are mutually independent Gaussian random variables. The channel inputs $X,X_1$ are real numbers and subject to the power constraints 
	\begin{align}
		\mathbb{E}(x^2)\le P, \
		\mathbb{E}(x_{1}^2)\le P_{1}.
	\end{align}
	The sensing parameter at the destination is $S_{d}=S_1$, and the quadratic distortion measure $d(s_{d},\hat{s}_{d})=(s_{d}-\hat{s}_{d})^2$ is considered. 
	
	In this example, the minimum distortion is $D_{\text{min}}=\frac{\sigma^2_{s_1}\sigma^2_{s_2}}{P_1+\sigma^2_{s_1}+\sigma^2_{s_2}}$. This can be achieved by setting $X=0$, $X_1\sim N(0,P_1)$, $V = S_1 + \tilde{S}_1$, where $\tilde{S}_1\sim N(0,\frac{\sigma^2_{s_1}\sigma^2_{s_2}}{P_1})$ is independent of $X,X_1,S_1,S_2$, in which the parameter estimation is performed based on $S_1 + \tilde{S}_1,S_1 + S_2$ actually.
	\begin{IEEEproof}
		The proof of optimality for $D_{\text{min}}=\frac{\sigma^2_{s_1}\sigma^2_{s_2}}{P_1+\sigma^2_{s_1}+\sigma^2_{s_2}}$ can be found in Appendix~\ref{appendix:proofExampleoptimalSensingGeneral}.
	\end{IEEEproof}
\end{example}

\par~In Example~\ref{example:optimalSensingGeneral}, the sensing parameter $S_{d}$ is correlated with the state $S_1$ that influences the channel outputs of both the relay and the destination. In this case, as shown in the discussions, the optimal strategy is letting the source sends $X=0$, the relay sends the compressed information $V=S_1+\tilde{S}_1$ by conducting $Y_1-X$ to obtain $S_1$ and choosing random variable $\tilde{S}_1$
and then the destination performs parameter estimation based on the compressed information $V$, its own received signal $Y$, and the decoded signals $X,X_1$. The compressed information $V$ here can be interpreted as the results of preprocessing at the relay to support the sensing estimation at the destination. 

\par~In what follows, we show that for two specific classes of relay channels, the relay can also adopt much simpler strategies to achieve the optimal sensing performance, i.e., $D_{\text{min}}$. These results provide fundamental insights for the investigation of the optimal capacity-distortion function $C(D)$ under specific channel conditions in Section~\ref{sec:optimalCD}.

\subsection{When the Sensing Parameter and the Channel to the Relay Are Independent}
\par~Consider the following class of relay channels in which the states that influence the channel outputs $Y,Y_1$ are mutually independent and sensing parameter is only correlated with the state that influences $Y$:
\begin{align}\label{class:relayChannelOptimalSensingDeterministic}
	\mathcal{C}_1 \triangleq \{\mathcal{X}\times\mathcal{X}_1,\mathcal{S}_1\times&\mathcal{S}_2\times\mathcal{S}_{d},\mathcal{Y}\times\mathcal{Y}_1,\notag\\
	&P_{S_1}P_{S_2S_{d}}P_{Y_1|XX_1S_1}P_{Y|XX_1S_2}\}.
\end{align}
For bistatic ISAC over any relay channels in $\mathcal{C}_1$, the relay obtains no information about $S_{d}$ from its observation $Y_1$, rendering the transmission of any compressed information $V$ to the destination superfluous. The minimum distortion is thus given by the following proposition. 

\begin{proposition}\label{proposition:relayChannelOptimalSensingDeterministic}
	For bistatic ISAC over any relay channels in $\mathcal{C}_1$, the minimum distortion is
	\begin{align}
		D_{\text{min}} = \min_{x,x_1} \mathbb{E}[d(S_{d},\hat{S}_{d})], 
	\end{align}
	where $\hat{S}_{d}$ is a deterministic function of $XX_1Y$ given as
	\begin{align}\label{equ:stateEstimatorSpecial}
		\hat{S}_{d}(x,x_1,y) = \operatorname*{argmin}_{s'_{d}\in\mathcal{S}_{d}}\sum_{s_{d}\in\mathcal{S}_{d}}P_{S_{d}|XX_1Y}(s_{d}|x,x_1,y)d(s_{d},s'_{d}).
	\end{align}
	\begin{IEEEproof}
		See Appendix~\ref{appendix:proofForPropositionrelayChannelOptimalSensingDeterministic}.
	\end{IEEEproof}
\end{proposition}

\begin{remark}\label{remark:optimalSensingInputCons}
	For bistatic ISAC over relay channels in $\mathcal{C}_1$, letting both the source and the relay send the deterministic signals enables the destination to achieve $D_{\text{min}}$. 
\end{remark}

\begin{example}\label{example:optimalSensingMAC}
	Consider a relay channel with binary inputs $X,X_1\in\{0,1\}$. The channel outputs at the relay and destination are
	\begin{align}
		Y_1 = S_1 X + S_2 X_1, \ Y = S_3X + S_4 X_1,
	\end{align} 
	where the channel state random variables $S_1,S_2,S_3,S_4$ are mutually independent binary random variables. The sensing state parameter at the destination is $S_{d}=S_3$, and the Hamming distortion measure $d(s_{d},\hat{s}_{d})=s_{d}\oplus\hat{s}_{d}$ is considered. 
	
	In this example, the minimum distortion is achieved by setting $X=1$ and $X_1=0$, i.e., both the source and the relay send deterministic signals, and the minimum distortion is $D_{\text{min}}=0$ by estimating the sensing parameter $S_3$ based on the $X,X_1,Y$.  
\end{example}

\subsection{When the Sensing Parameter and the Channel to the Destination Are Independent}
\par~Consider the following class of relay channels in which the states that influence the channel outputs $Y,Y_1$ are mutually independent and sensing parameter is only correlated with the state that influences $Y_1$:
\begin{align}\label{class:relayChannelOptimalSensingEstimatedForward}
	\mathcal{C}_2 \triangleq \{\mathcal{X}\times\mathcal{X}_1,\mathcal{S}_1\times&\mathcal{S}_2\times\mathcal{S}_{d},\mathcal{Y}\times\mathcal{Y}_1,\notag\\
	&P_{S_1S_{d}}P_{S_2}P_{Y_1|XX_1S_1}P_{Y|XX_1S_2}\}.
\end{align}
For bistatic ISAC over any relay channels in $\mathcal{C}_2$, the destination obtains no information about $S_{d}$ from its observation $Y$ except that the relay sends compressed information about $S_{d}$. The minimum distortion is thus given by the following proposition. 
\begin{proposition}\label{proposition:OptimalSensingEstimateForward}
	For bistatic sensing over any relay channels in $\mathcal{C}_2$, the minimum distortion is 
	\begin{align}
		D_{\text{min}} = \min_{P_{XX_1}P_{\hat{S}_{d}|XX_1Y_1}} \mathbb{E}[d(S_{d},\hat{S}_{d})], 
	\end{align}
	where the minimum is over all pmfs $P_{XX_1}P_{\hat{S}_{d}|XX_1Y_1}$ such that 
	\begin{align}\label{equ:consProposition2}
		I(X_1;Y|X)\ge I(\hat{S}_{d};Y_1|XX_1).
	\end{align}
	and the joint distribution of variables $XX_1S_1S_2S_{d}YY_1\hat{S}_{d}$ is 
	\begin{align}
		P_{XX_1}P_{S_1S_{d}}P_{S_2}P_{Y_1|XX_1S_1}P_{Y|XX_1S_2}P_{\hat{S}_{d}|XX_1Y_1}. 
	\end{align}
	\begin{IEEEproof}
		See Appendix~\ref{appendix:ProofForPropositionOptimalSensingEstimateForward}.
	\end{IEEEproof}
\end{proposition}

\begin{remark}
	For bistatic ISAC over relay channels in $\mathcal{C}_2$, to achieve the minimum distortion, the relay can perform the estimation $\hat{S}_{d}$ firstly based on the received signal $Y_1$, its own transmitted signal $X_1$, and the pilot signal $X$ sent by the source. Such an estimation $\hat{S}_{d}$ is then forwarded to the destination as the parameter estimation result. The optimization of the channel input distribution $P_{XX_1}$ is therefore a result that takes into account both parameter estimation $\hat{S}_{d}$ and data transmission from the relay to the destination.
\end{remark}

\begin{example}\label{example:optimalSensingEstimatingForward}
	Consider a Gaussian relay channel with additive Gaussian states
	\begin{align}
		Y_1 = X + S_1, \ Y = X + X_1 + S_2,
	\end{align}
	where $S_1\sim N(0,\sigma^2_{s_1})$ and $S_2\sim N(0,\sigma^2_{s_2})$ are mutually independent, and the channel inputs $X,X_1$ are real numbers and subject to the power constraints 
	\begin{align}
		\mathbb{E}(x^2)\le P, \ 
		\mathbb{E}(x_{1}^2)\le P_{\text{1}}.
	\end{align}
	The sensing parameter at the destination is $S_{d}=S_1$, and the quadratic distortion measure $d(s_{d},\hat{s}_{d})=(s_{d}-\hat{s}_{d})^2$ is considered. 
	
	In this example, the minimum distortion is $D_{\text{min}}=\frac{\sigma^2_{s_1}\sigma^2_{s_2}}{P_1+\sigma^2_{s_2}}$. This can be achieved by setting $X=0,X_1\sim N(0,P_1),\hat{S}_1=\tilde{S}_1$, where $S_1=\tilde{S}_1+E$, and $E\sim N(0,\frac{\sigma^2_{s_1}\sigma^2_{s_2}}{P_1+\sigma^2_{s_2}})$ is independent of $X,X_1,S_1,S_2,\tilde{S}_1$.  
	\begin{IEEEproof}
		The proof of optimality can be found in Appendix~\ref{appendix:proofOFExample3}.
	\end{IEEEproof}
\end{example}

\section{Optimal Capacity-Distortion Functions for Specific Bistatic ISAC Relay Channels}\label{sec:optimalCD}
\par~In this section, we provide three specific classes of relay channels where the established upper and lower bounds presented in Section~\ref{sec:outerAndInner} coincide, and thus leading to optimal capacity-distortion function $C(D)$. In particular, we show that for the considered three classes of relay channels, partial-decode-forward strategy, compress-forward strategy, and hybrid-partial-decode-and-compress-forward strategy achieve the optimal capacity-distortion functions, respectively.

\subsection{Relay Channels with Orthogonal Sender Components: Optimal Tradeoff Achieved by Partial-Decode-Forward Strategy}
\par
\begin{figure}[!t]
	\centering
	\includegraphics[width=.99\linewidth]{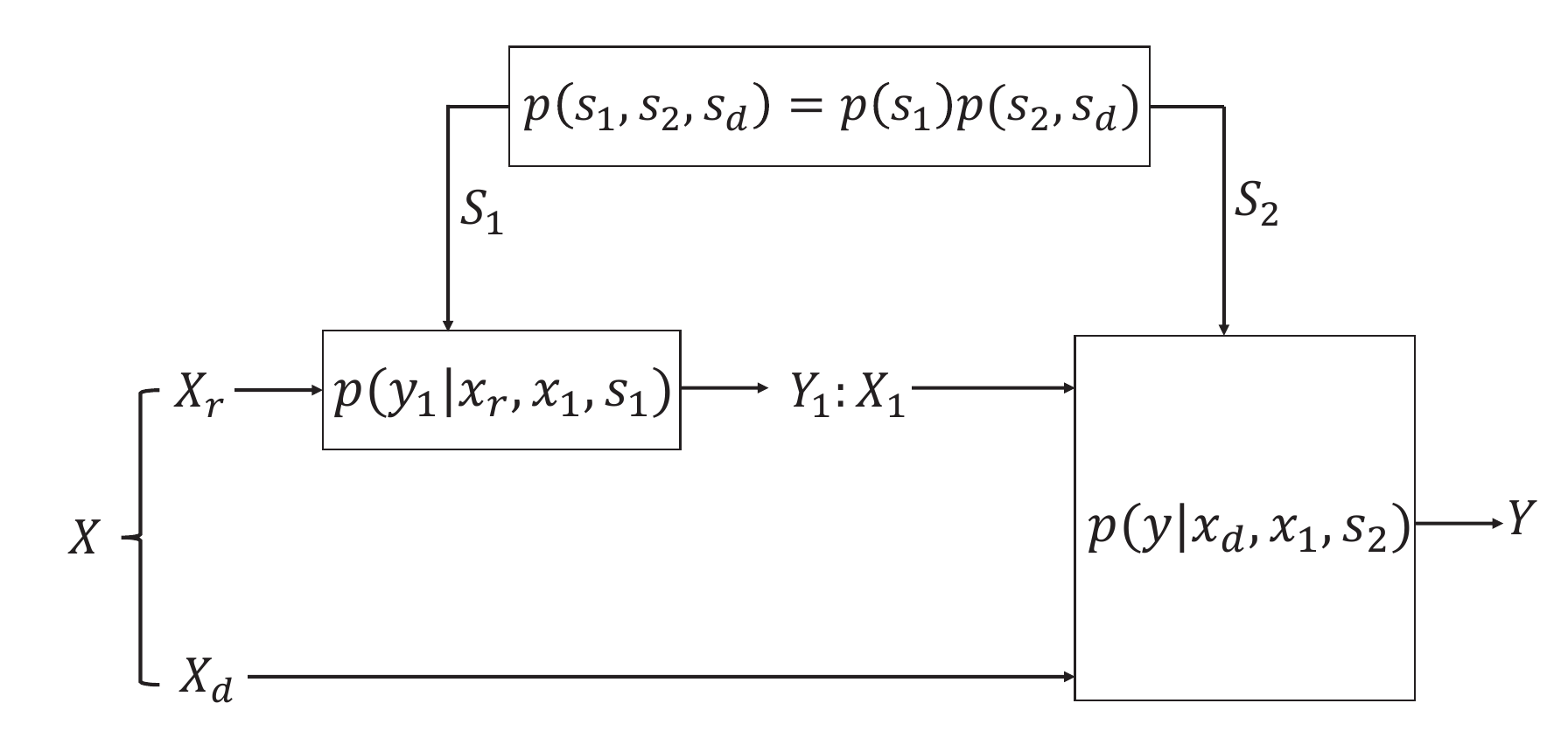}
	\caption{Bistatic ISAC over relay channels with orthogonal sender components $X=(X_r,X_d)$, independent states $S=(S_1,S_2)$, and sensing parameter $S_d$ correlated with $S_2$.}
	\label{fig:RC_Orthogonal_Independent_inputCons}
\end{figure}

~Consider the following class of bistatic ISAC relay channels: 
\begin{align}\label{equ:channelSpecialOrthogonalInputCons}
	\mathcal{C}_3 \triangleq\{\mathcal{X}_r\times\mathcal{X}_d\times&\mathcal{X}_1, \mathcal{S}_1\times \mathcal{S}_2\times \mathcal{S}_d,\mathcal{Y}\times\mathcal{Y}_1:\notag\\
	&P_{S_1}P_{S_2S_d}P_{Y_1|X_rX_1S_1}P_{Y|X_dX_1S_2}
	\},
\end{align}
which is illustrated in Fig.~\ref{fig:RC_Orthogonal_Independent_inputCons}.
A representative scenario corresponding to the channel model in $\mathcal{C}_3$ is a connected vehicular network, as illustrated in Fig.~\ref{fig:motivating}. In this setting, the signals transmitted from the base station to the relay and the destination are orthogonal, for example, implemented through frequency-division multiplexing. The sensing targets in this scenario act as scattering objects that influence the received signals at the destination.
The class of relay channels $\mathcal{C}_3$ is a special case of $\mathcal{C}_1$. Therefore, relay obtains no useful information for sensing at the destination. Moreover, by leveraging the findings in~\cite{el2005capacity} to relay channels in $\mathcal{C}_3$, we can establish that the partial-decode-forward strategy attains the communication capacity. Combined with these facts, it can be proved that the partial-decode-forward strategy achieves the optimal capacity-distortion tradeoff $C(D)$ for bistatic ISAC over relay channels in $\mathcal{C}_3$. 

\begin{theorem}\label{proposition:RC_orthogonal_independent_inputCons}
	The capacity-distortion function $C(D)$ of bistatic ISAC over the class of relay channels in ${\mathcal C}_3$ is given by
	\begin{align}
		C(D) = \max&_{P_{X_1}P_{X_r|X_1}P_{X_d|X_1}\in\mathcal{P}_D} \min\bigg\{I(X_dX_1;Y),\notag\\
		&\qquad I(X_r;Y_1|X_1)+I(X_d;Y|X_1)\bigg\},
	\end{align}
	where
	\begin{align}
		\mathcal{P}_D = \bigg\{P_{X_1}P_{X_r|X_1}P_{X_d|X_1}:\mathbb{E}[d(S_d,\hat{S}_d)]\le D~\bigg\},
	\end{align}
	$\hat{S}_d$ is a deterministic function of $X_d,X_1,Y$ given as
	\begin{align}
		\hat{S}_{d}(&x_d,x_1,y)=\notag\\
		& \operatorname*{argmin}_{s'_d\in\mathcal{S}_d}\sum_{s_d\in\mathcal{S}_d}P_{S_d|X_dX_1Y}(s_d|x_d,x_1,y)d(s_d,s'_d), 
	\end{align}
	and the joint distribution of $X_rX_dX_1S_1S_2S_dYY_1\hat{S}_d$ is 
	\begin{align}
		P_{X_1}P_{X_r|X_1}P_{X_d|X_1}P_{S_1}P_{S_2S_d}&P_{Y_1|X_rX_1S_1}\notag\\
		&P_{Y|X_dX_1S_2}P_{\hat{S}_d|X_dX_1Y}. 
	\end{align}
	\begin{IEEEproof}
		The proof follows directly from that of Proposition~\ref{proposition:relayChannelOptimalSensingDeterministic} combined with the proof techniques for relay channels with orthogonal sender components presented in~\cite{el2005capacity}, and is thus omitted here. 
	\end{IEEEproof}
\end{theorem}

\begin{remark}
	For bistatic ISAC over relay channels in ${\mathcal C}_3$, the optimal capacity-distortion function $C(D)$ is achieved by partial-decode-forward strategy, and the parameter estimation is simply based on the received signal $Y$ and decoded transmitted signals $X_d,X_1$. The distortion constraint for the parameter estimation is thus reduced to an additional cost constraint on the input distribution $P_{X_1}P_{X_r|X_1}P_{X_d|X_1}$.
\end{remark} 

\begin{remark}
	When the channel transition probability degrades into $P_{Y_1|X_rS_1}P_{Y|X_1S_2}$, the relay channel in ${\mathcal C}_3$ specializes into a two-hop line channel, and our results in Theorem~\ref{proposition:RC_orthogonal_independent_inputCons} recover those detailed in~\cite{salimi2017capacity} for sensing estimation $S_{d}=S_2$ as a special case.
\end{remark} 

 \begin{example}\label{example:optimalOrthogonalInputCons}
	Consider a relay channel with orthogonal sender components, i.e., $X=(X_r,X_d)$. The channel inputs $X_r,X_d,X_1$ are binary. The channel output at the relay is $Y_1=S_1X_r$, and the channel output at the destination is $Y=S_2X_d+S_3X_1$, where the states $S_1,S_2,S_3$ are mutually independent binary random variables. The sensing state at the destination is $S_d=S_2$, and the Hamming distortion $d(s_d,\hat{s}_d)=s_d\oplus\hat{s}_d$ is considered. 
	
	\begin{figure}[!t]
		\centering
		\includegraphics[width=.96\linewidth]{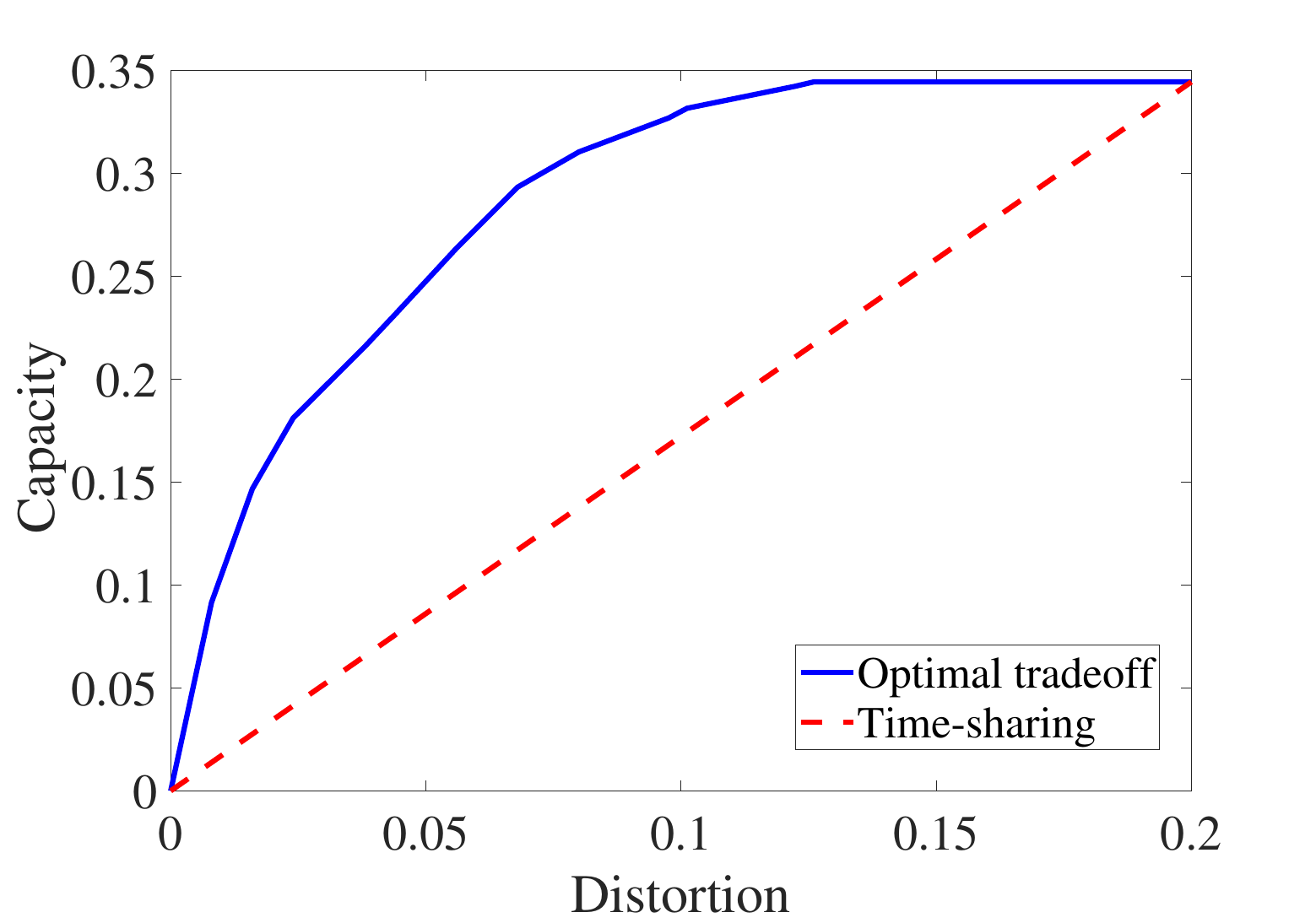}
		\caption{Capacity-distortion tradeoff $C(D)$ for the channel considered in Example~\ref{example:optimalOrthogonalInputCons}.}
		\label{fig:optimalOrthogonalInputCons}
	\end{figure}
	This channel is an instance of the relay channels within ${\mathcal C}_3$ as defined in~\eqref{equ:channelSpecialOrthogonalInputCons}. Thus, the optimal capacity-distortion tradeoff $C(D)$ is given in Theorem~\ref{proposition:RC_orthogonal_independent_inputCons}. Let $a\triangleq P_{X_1}(1)$, $b\triangleq P_{X_r|X_1}(1|1)$, $c\triangleq P_{X_r|X_1}(1|0)$, $d\triangleq P_{X_d|X_1}(1|1)$, $e\triangleq P_{X_d|X_1}(1|0)$.  The optimal capacity-distortion function $C(D)$ can be evaluated as
	\begin{subequations}
		\begin{align}
			C(&D) = \max_{a,b,c,d,e}\min\bigg\{\alpha,\beta\bigg\}, \\
			&\text{s.t.}\quad a,b,c,d,e\in[0,1], \\
			&\bigg((1-a)(1-e)+a(1-d)\bigg)\min\{P_{S_2},1-P_{S_2}\}\notag\\
			&\quad+ ad\min\{(1-P_{S_2})P_{S_3},P_{S_2}(1-P_{S_3})\} \le D,
		\end{align}
	\end{subequations}
	where 
	\begin{subequations}
		\begin{align}
			\alpha =& H_3\bigg(adP_{S_2}P_{S_3},ad(P_{S_2}*P_{S_3})+ (1-a)eP_{S_2}\notag\\
			&+a(1-d)P_{S_3}\bigg)
			-adH_3(P_{S_2}P_{S_3},P_{S_2}*P_{S_3}) \notag\\
			&- (1-a)eH_2(P_{S_2})-a(1-d)H_2(P_{S_3}),\\
			\beta=&(1-a)H_2(cP_{S_1}) -\bigg(ab+(1-a)c\bigg)H_2(P_{S_1})\notag\\
			&+ aH_2(bP_{S_1})+aH_3\bigg(dP_{S_2}P_{S_3},(1-P_{S_3})(1-dP_{S_2})\bigg)\notag\\
			&+(1-a)H_2(eP_{S_2})-adH_3(P_{S_2}P_{S_3},P_{S_2}*P_{S_3})\notag\\
			&-(1-a)eH_2(P_{S_2})-a(1-d)H_2(P_{S_3}).
		\end{align}
	\end{subequations}

	Given $P_{S_1} =0.4$, $P_{S_2}=0.2$, and $P_{S_3}=0.6$, the capacity-distortion function $C(D)$ is as shown in Fig.~\ref{fig:optimalOrthogonalInputCons}. Two extreme cases are considered:
	\begin{itemize}
		\item Optimal sensing, achieved by setting $(X_d,X_1)=(1,0)$, enables the destination to perfectly acquire $S_2$, resulting in minimal distortion $D_{\text{min}}=0$ and zero communication rate.
		\item Optimal communication, without the distortion constraint, yields communication capacity $C_{\text{max}}=0.3444$.
	\end{itemize}
	We also examine the time-sharing approach that involves alternating between pure-sensing and pure-communication operation points, specifically $(D_{\text{min}}=0,0)$ and $(0.2,C_{\text{max}}=0.3444)$. As shown in Fig.~\ref{fig:optimalOrthogonalInputCons}, the unified design provides a substantial gain over the time-sharing approach. 
\end{example}

\subsection{State-Dependent Cover-Kim Relay Channels: Optimal Tradeoff Achieved by Compress-Forward Strategy}
\par~Consider the following class of bistatic ISAC relay channels: 
\begin{align}\label{equ:CoverKimRelayChannel}
	\mathcal{C}_4 \triangleq& \{\mathcal{X}\times\mathcal{X}_1, \mathcal{S}\times \mathcal{S}_d,\mathcal{Y}_r\times\mathcal{Y}_d\times\mathcal{Y}_1:\notag\\
	&P_{SS_d}P_{Y_dY_1|XS}P_{Y_r|X_1},~\text{and}~Y_1=f(X,Y_d)
	\},
\end{align}
where $f(\cdot)$ is a deterministic function. 
A representative scenario corresponding to the channel model in $\mathcal{C}_4$ is a special case of connected vehicular network. In this scenario, the relay is a small base station connected to a macro base station (source) via a low-noise and amplifier-free optical fiber. Thus, the sample signal at the relay can be regarded, to within digital-resolution accuracy, as a deterministic function of transmitted signals $X$ at the source.  The relay using orthogonal frequency band to communicate with the vehicular node (destination), and the sensing targets act as scattering objects that influence the received signals at the destination.
The channels in $\mathcal{C}_4$ are the Cover-Kim relay channels~\cite{kim2008capacity}. Thus, the compress-forward strategy attains the communication capacity\footnote{The hash-forward strategy can also achieve the communication capacity. In this paper, we focus on the compress-forward strategy.}. It can be further proved that the compress-forward strategy achieves the optimal capacity-distortion tradeoff $C(D)$ for bistatic ISAC over relay channels in $\mathcal{C}_4$.

\begin{theorem}\label{theorem:cover-kimRelayChannel}
	The capacity-distortion function $C(D)$ of bistatic ISAC over the class of relay channels in ${\mathcal C}_4$ is given by
	\begin{align}
		C(D) = \max_{P_{X}P_{X_1}\in\mathcal{P}_D}~&\min\bigg\{I(X;Y_1Y_d),\notag\\
		&I(X_1;Y_r)+I(X;Y_d)\bigg\},
	\end{align}
	where 
	\begin{align}
		\mathcal{P}_D = \bigg\{P_{X}P_{X_1}:\mathbb{E}[d(S_d,\hat{S}_d)]\le D\bigg\},
	\end{align}
	$\hat{S}_d$ is a deterministic function of $X,Y_d$ given as
	\begin{align}
		\hat{S}_{d}(x,y_d)& = \operatorname*{argmin}_{s'_d\in\mathcal{S}_d}\sum_{s_d\in\mathcal{S}_d}P_{S_d|XY_d}(s_d|x,y_d)d(s_d,s'_d), 
	\end{align}
	and the joint distribution of $XX_1SS_dY_rY_dY_1\hat{S}_d$ is 
	\begin{align}
		P_{X}P_{X_1}P_{SS_d}P_{Y_dY_1|XS}P_{Y_r|X_1}P_{\hat{S}_d|XY_d}. 
	\end{align}
	\begin{IEEEproof}
		See Appendix~\ref{appendix:cover-kimRelayChannel}.
	\end{IEEEproof}
\end{theorem}

\begin{remark}
	For bistatic ISAC over relay channels in ${\mathcal C}_4$, the optimal capacity-distortion function $C(D)$ can be achieved by letting the relay send the received signal $Y_1$ as the compressed information by incorporating the $X,X_1,Y_r,Y_d$ as the side information. The parameter estimation is simply based on the received signal $Y_d$ and decoded transmitted signal $X$, which thus makes the distortion constraints for the parameter estimation become an additional cost constraint on the input distribution $P_{X}P_{X_1}$.
\end{remark}

\begin{figure}[!t]
	\centering
	\includegraphics[width=.99\linewidth]{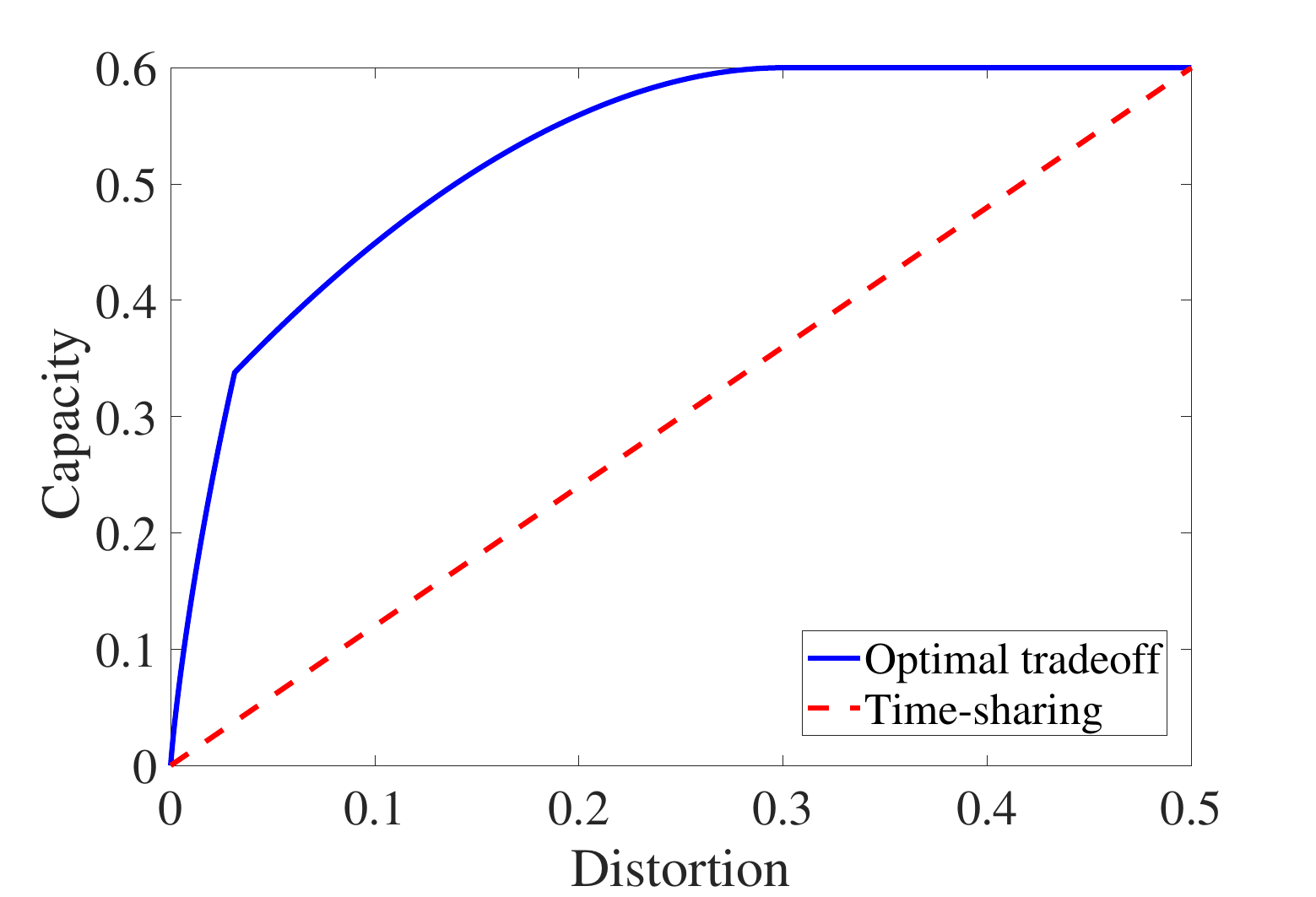}
	\caption{Capacity-distortion tradeoff $C(D)$ for the channel considered in Example~\ref{example:coverKimRelayChannel}.}
	\label{fig:coverKimRelay}
\end{figure}

\begin{example}\label{example:coverKimRelayChannel}
	Consider a relay channel with binary inputs $X,X_1\in\{0,1\}$ and orthogonal receiver components $Y=(Y_r,Y_d)$. The channel output at the relay is $Y_1=(S+1)X$, and the channel output at the destination is $Y_r=X_1\oplus N, Y_d=SX$, where the states $S,N$ are mutually independent binary random variables. The sensing state at the destination is $S_d=S$, and the Hamming distortion $d(s_d,\hat{s}_d)=s_d\oplus\hat{s}_d$ is considered.
	
	This channel is an instance of the relay channels within ${\mathcal C}_4$ as defined in~\eqref{equ:CoverKimRelayChannel}. Thus, the optimal capacity-distortion function $C(D)$ is given in Theorem~\ref{theorem:cover-kimRelayChannel}. Let $a\triangleq P_{X}(1)$, $b\triangleq P_{X_1}(1)$. The optimal capacity-distortion function $C(D)$ can be evaluated as 
	\begin{subequations}
		\begin{align}
			&C(D) = \max_{a,b}~ \min\bigg\{H_2(a),\notag\\
			&\quad H_2(b*P_N)-H_2(P_N)+H_2(aP_s)-aH_2(P_S)\bigg\}, \\
			&\text{s.t.}\quad a,b\in[0,1], \\
			&\qquad (1-a)\min\{P_S,1-P_S\}\le D.
		\end{align}
	\end{subequations}
	
	Given $P_S=0.5$ and $P_N=0.2$, the capacity-distortion function $C(D)$ is as shown in Fig.~\ref{fig:coverKimRelay}. Two extreme cases are considered:
	\begin{itemize}
		\item Optimal sensing, achieved by setting $X=1$, enables the destination to perfectly acquire $S$, resulting in minimal distortion $D_{\text{min}}=0$ and zero communication rate.
		\item Optimal communication, without the distortion constraint, yields communication capacity $C_{\text{max}}=0.6$.
	\end{itemize}
	We also examine the time-sharing approach that involves alternating between pure-sensing and pure-communication operation points, specifically $(D_{\text{min}}=0,0)$ and $(0.5,C_{\text{max}}=0.6)$. As shown in Fig.~\ref{fig:coverKimRelay}, the unified design provides a substantial gain over the time-sharing approach.
\end{example}

\subsection{Two-hop Relay Channels: Optimal Tradeoff Achieved by Hybrid-Partial-Decode-and-Compress-Forward Strategy}
\par
\begin{figure}[!t]
	\centering
	\includegraphics[width=.99\linewidth]{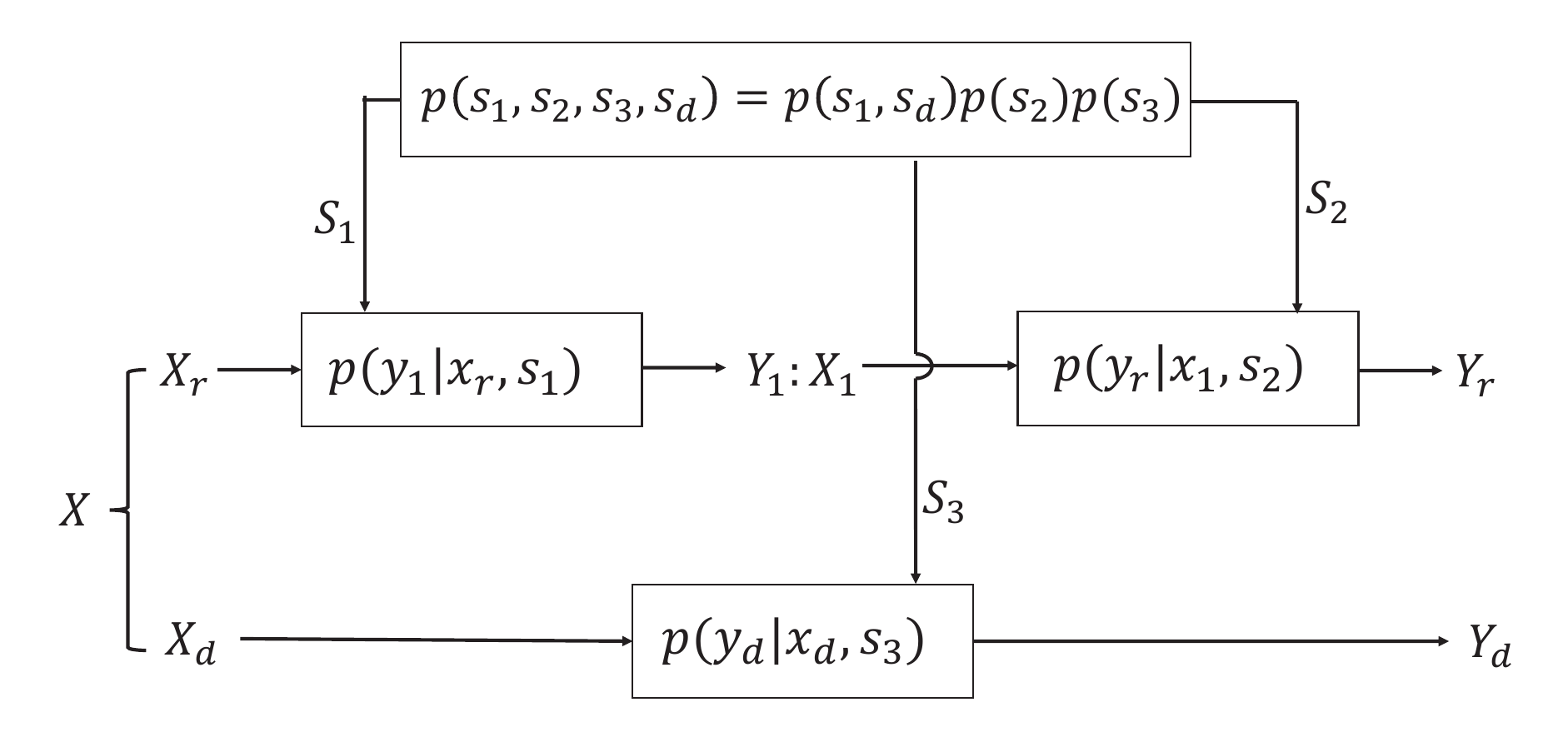}
	\vspace{-1.5em}
	\caption{Bistatic ISAC over relay channels with orthogonal sender components $X=(X_r,X_d)$, orthogonal receiver components $Y=(Y_r,Y_d)$, independent states $S=(S_1,S_2,S_3)$, and sensing state $S_d$ correlated with $S_1$.}
	\label{fig:RC_Orthogonal_estimatedForward}
\end{figure}

~Consider the following class of bistatic ISAC relay channels: 
\begin{align}\label{equ:channelSpecialOrthogonalEstimatedForward}
	\mathcal{C}_5 \triangleq& \{\mathcal{X}_r\times\mathcal{X}_d\times\mathcal{X}_1, \mathcal{S}_1\times \mathcal{S}_2\times \mathcal{S}_3\times \mathcal{S}_d,\mathcal{Y}_r\times\mathcal{Y}_d\times\mathcal{Y}_1:\notag\\
	&P_{S_1S_d}P_{S_2}P_{S_3}P_{Y_1|X_rS_1}P_{Y_r|X_1S_2}P_{Y_d|X_dS_3}
	\},
\end{align}
which is illustrated in Fig.~\ref{fig:RC_Orthogonal_estimatedForward}. A representative scenario corresponding to the channel model in $\mathcal{C}_5$ is also a connected vehicular network, as illustrated in Fig.~\ref{fig:motivating}. In this scenario, the transmitted signals from all nodes to their neighboring nodes are orthogonal, for example, implemented through frequency-division multiplexing. The sensing targets act as scattering objects that influence the received signals at the relay. The class of relay channels $\mathcal{C}_5$ is a special case of $\mathcal{C}_2$. Therefore, the sensing at the destination entirely depends on the compressed information $V$ sent by the relay. Moreover, it can be found that partial-decode-forward strategy attains the communication capacity for the channels in $\mathcal{C}_5$. Combined with these facts, it can be proved that the hybrid-partial-decode-and-compress-forward strategy achieves the optimal capacity-distortion tradeoff $C(D)$ for bistatic ISAC over relay channels in $\mathcal{C}_5$.

\begin{theorem}\label{proposition:RC_orthogonal_hybrid_decode_compress_forward}
	The capacity-distortion function $C(D)$ of bistatic ISAC over the class of relay channels in ${\mathcal C}_5$ is given by
	\begin{align}
		C(D) &= \max_{P_{X_r}P_{X_d}P_{X_1}P_{\hat{S}_d|X_rY_1}\in\mathcal{P}_D}~ I(X_d;Y_d) \notag\\
		&+ \min\bigg\{I(X_r;Y_1),I(X_1;Y_r)-I(\hat{S}_d;Y_1|X_r)\bigg\},
	\end{align}
	where 
	\begin{align}
		\mathcal{P}_D = \bigg\{&P_{X_r}P_{X_d}P_{X_1}P_{\hat{S}_d|X_rY_1}:\mathbb{E}[d(S_d,\hat{S}_d)]\le D~\notag\\
		&\qquad\qquad\quad\text{and}~I(X_1;Y_r)\ge I(\hat{S}_d;Y_1|X_r)\bigg\}.
	\end{align}
	Note that $\hat{S}_d$ is not necessary to be a deterministic function of $X_r,Y_1$, and the joint distribution of variables $X_rX_dX_1S_1S_2$ $S_3S_dY_rY_dY_1\hat{S}_d$ is 
	\begin{align}
		P_{X_r}P_{X_d}P_{X_1}P_{S_1S_d}P_{S_2}P_{S_3}P_{Y_1|X_rS_1}&P_{Y_r|X_1S_2}\notag\\
		&P_{Y_d|X_dS_3}P_{\hat{S}_d|X_rY_1}. 
	\end{align}
	\begin{IEEEproof}
		The channels in $\mathcal{C}_5$ are combinations of a two-hop line channel~\cite{salimi2017capacity} and a direct channel from destination to the source. The capacity-distortion function is thus an extension of the results in~\cite{salimi2017capacity} by introducing an additional mutual information term $I(X_d;Y_d)$ in capacity-distortion function $C(D)$. The proof is thus omitted here. 
	\end{IEEEproof}
\end{theorem}

\begin{remark}
	For bistatic ISAC over relay channels in ${\mathcal C}_5$, the optimal capacity-distortion function $C(D)$ is achieved by hybrid-partial-decode-and-compress-forward strategy. In particular, the relay decodes partial message sent by the source and perform the parameter estimation based on the received signal. Then, it sends both the decoded message and estimated parameter as the compressed information to the destination for bistatic ISAC.
\end{remark}

\begin{example}\label{example:optimalOrthogonalInputOrthogonalOutput}
	Consider a relay channel with orthogonal sender components $X=(X_r,X_d)$ and orthogonal receiver components $Y=(Y_r,Y_d)$. The channel inputs $X_r,X_d,X_1$ are binary. The channel output at the relay is $Y_1=S_1X_r$, and the channel output at the destination is $Y_r=S_2X_1,Y_d=S_3X_d$, where the states $S_1,S_2,S_3$ are mutually independent binary random variables. The sensing state at the destination is $S_d=S_1$, and the Hamming distortion $d(s_d,\hat{s}_d)=s_d\oplus\hat{s}_d$ is considered. 
	
	This channel is an instance of the relay channels within ${\mathcal C}_5$ as defined in~\eqref{equ:channelSpecialOrthogonalEstimatedForward}. Thus, the optimal capacity-distortion function $C(D)$ is given in Theorem~\ref{proposition:RC_orthogonal_hybrid_decode_compress_forward}. Let $a\triangleq P_{X_r}(1)$, $b\triangleq P_{X_d}(1)$, $c\triangleq P_{X_1}(1)$, $d\triangleq P_{\hat{S}_d|X_rY_1}(1|0,0)$, $e\triangleq P_{\hat{S}_d|X_rY_1}(1|1,0)$, $f\triangleq P_{\hat{S}_d|X_rY_1}(1|1,1)$. The optimal capacity-distortion function $C(D)$ can be evaluated as 
	\begin{subequations}
		\begin{align}
			C(&D) = \max_{a,b,c,d,e,f}~ \alpha + \min\bigg\{\beta, \gamma-\eta\bigg\}, \\
			&\text{s.t.}\quad a,b,c,d,e,f\in[0,1], \\
			&\qquad \gamma \ge \eta, \\
			&\qquad (1-a)\bigg((1-d)P_{S_1}+d(1-P_{S_1})\bigg)\notag\\
			&\qquad\qquad\quad+ae(1-P_{S_1})+a(1-f)P_{S_1}\le D,
		\end{align}
	\end{subequations}
	where 
	\begin{subequations}
		\begin{align}
			\alpha =& H_2(bP_{S_3})-bH_2(P_{S_3}) ,\\
			\beta=&H_2(aP_{S_1})-aH_2(P_{S_1}), \\
			\gamma = &H_2(cP_{S_2})-cH_2(P_{S_2}),\\
			\eta =& aH_2\bigg(e(1-P_{S_1})+fP_{S_1})\bigg)\notag\\
			&\qquad-a(1-P_{S_1})H_2(e)-aP_{S_1}H_2(f).
		\end{align}
	\end{subequations}
	
	\begin{figure}[!t]
		\centering
		\includegraphics[width=.99\linewidth]{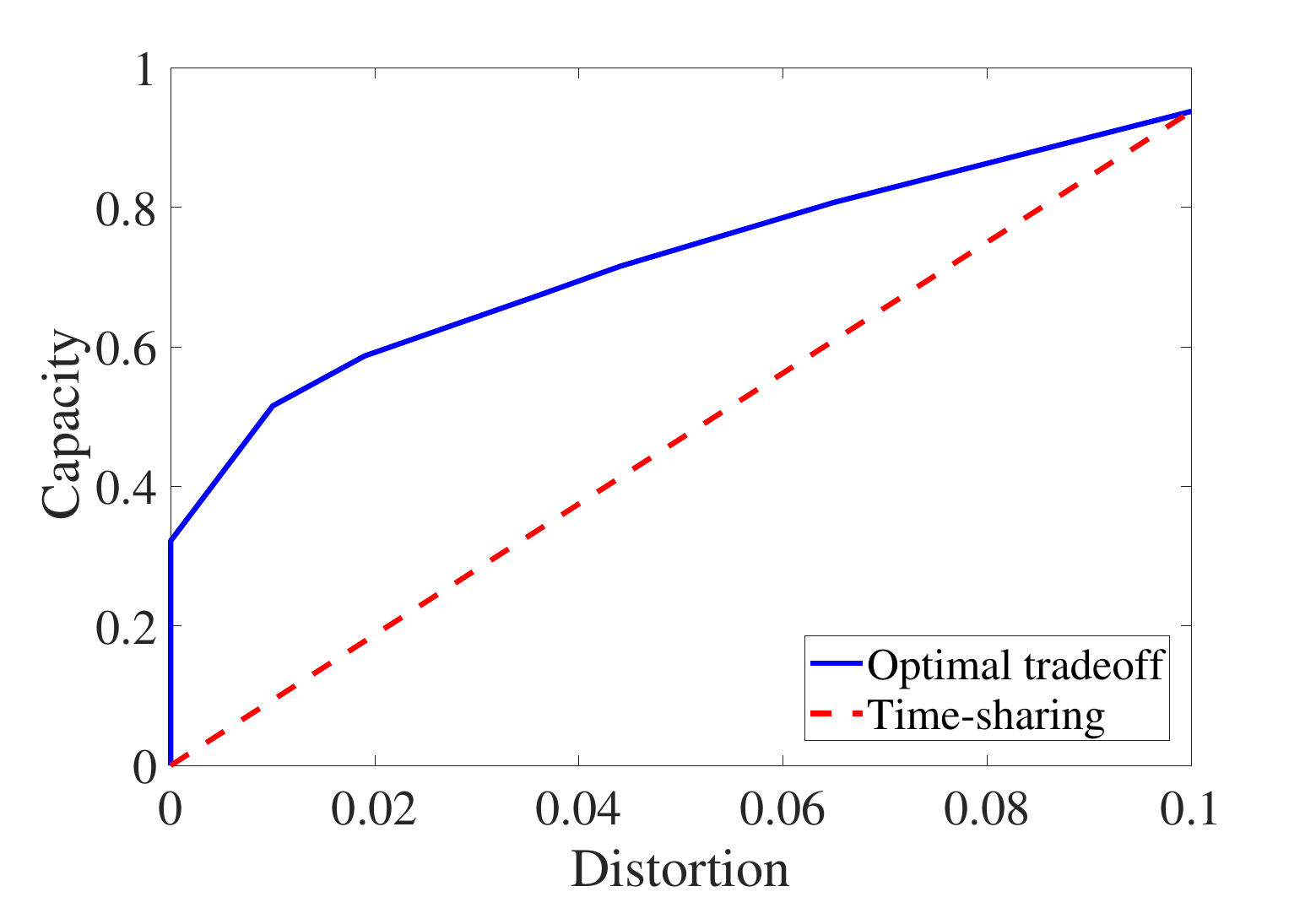}
		\vspace{-1.5em}
		\caption{Capacity-distortion tradeoff $C(D)$ for the channel considered in Example~\ref{example:optimalOrthogonalInputOrthogonalOutput}.}
		\label{fig:optimalHybridDecodeCompress}
	\end{figure}
	Given $P_{S_1} =0.9$, $P_{S_2}=0.8$, and $P_{S_3}=0.5$, the capacity-distortion function $C(D)$ is as shown in Fig.~\ref{fig:optimalHybridDecodeCompress}. Two extreme cases are considered:
	\begin{itemize}
		\item Optimal sensing, achieved by setting $X_r=1$, enables the relay to perfectly acquire $S_2$ and transmit it to the destination, resulting in minimal distortion $D_{\text{min}}=0$ and communication capacity $C=0.3219$, where the communication is achieved through the direct channel from the source to the destination with transition probability $Y_d=S_3X_d$.
		\item Optimal communication, without the distortion constraint, yields communication capacity $C_{\text{max}}=0.9375$.
	\end{itemize}
	We also examine the time-sharing approach that involves alternating between pure-sensing and pure-communication operation points, specifically $(D_{\text{min}}=0,0)$ and $(0.1,C_{\text{max}}=0.9375)$. As shown in Fig.~\ref{fig:optimalHybridDecodeCompress}, the unified design provides a substantial gain over the time-sharing approach. 
\end{example}

\section{Conclusion}\label{sec:conclusion}
\par~In this paper, we have investigated the fundamental limits of bistatic ISAC over discrete memoryless relay channels. Both upper and lower bounds on the capacity-distortion function have been established. We have further proved that the proposed scheme achieves optimal sensing performance, i.e., minimum distortion for sensing, when the communication task is ignored. We have also identified the optimal capacity-distortion tradeoff for three specific classes of relay channels by enabling the relay to choose different strategies. While the upper and lower bounds presented in this paper coincide with each other for specific cases, the optimal capacity-distortion tradeoff for bistatic ISAC over the general relay channels remains open. 

\par~Finally, we note that both our current work and the prior studies~\cite{kobayashi2018joint,liu2022information,kobayashi2019joint,ahmadipour2023information,liu2023Globecom,liu2022generalized,ahmadipour2022information,liu2024ICC,xiong2023fundamental,gunlu2023secure,welling2024transmitter,nikbakht2024integrated,zhang2011joint,salimi2017capacity,jiao2024rate,ahmadipour2023strong} focus on the memoryless ISAC channels with i.i.d. states. This modeling framework is widely adopted to capture the fast time-varying characteristics of ISAC channel parameters. Under such assumptions, existing studies have successfully established single-letter characterizations of the communication-sensing performance tradeoff and derived several optimal results under general or specific channel conditions. In more practical scenarios, however, channels often exhibit memory, where past channel echos and state estimates influence current and future channel outputs and estimates. Under this conditions, obtaining a single-letter characterization of the performance tradeoff becomes intractable, and evaluating the tradeoff itself poses significant analytical challenges. Recent studides~\cite{chen2023general,nikbakht2024memory,lindstrom2025rate} have begun to  address point-to-point ISAC channels with memory and have derived multi-letter expressions for the capacity-distortion tradeoff under various channel conditions. Nonetheless, evaluating these expressions remains difficult, as it requires solving complex multi-letter joint distribution optimization problems. Although~\cite{nikbakht2024memory} proposed a preliminary reinforcement learning-based approach using the deep deterministic policy gradient algorithm, the exact optimal tradeoff still remains unknown. Looking forward, it is of great importance to explore new analytical and computational tools, such as neural estimation framework~\cite{belghazi2018mutual}, to more explicitly evaluate the performance tradeoff. Moreover, extending the study to investigate the fundamental limits of general multi-terminal ISAC networks with channel memory and finite blocklength constraints represents a key direction for future research. 

\appendices
\section{Proof of Theorem~\ref{theorem:outer}}\label{appendix:proofOuter}
\par~We identify the auxiliary random variables $T_i \triangleq WY_{i+1}^nY_{1}^{i-1},~i\in[1:n]$ with $Y_{1,0}=Y_{n+1}=\phi$. Note that, $T_i-X_iX_{1,i}Y_{1,i}-Y_iS_iS_{d,i}$ forms a Markov chain for $i\in[1:n]$. We have
\begin{align}\label{equ:rateBoundKeyOne}
	&nR \overset{(a)}\le I(W;Y^n) + n\epsilon_n \notag\\
	& = \sum_{i=1}^n I(W;Y_i|Y_{i+1}^n) + n\epsilon_n \notag\\
	& \le \sum_{i=1}^n I(WY_{i+1}^n;Y_i) + n\epsilon_n \notag\\
	& = \sum_{i=1}^n I(WY_{i+1}^nY_{1}^{i-1};Y_i) - \sum_{i=1}^nI(Y_{1}^{i-1};Y_i|WY_{i+1}^n)+ n\epsilon_n \notag\\
	& \overset{(b)} =  \sum_{i=1}^n I(WY_{i+1}^nY_{1}^{i-1};Y_i) - \sum_{i=1}^nI(Y_{i+1}^{n};Y_{1,i}|WY_{1}^{i-1})+ n\epsilon_n \notag\\
	& \overset{(c)} = \sum_{i=1}^n I(WY_{i+1}^nY_{1}^{i-1}X_iX_{1,i};Y_i)\notag\\
	&\qquad- \sum_{i=1}^nI(Y_{i+1}^{n};Y_{1,i}|WY_{1}^{i-1}X_iX_{1,i})+ n\epsilon_n \notag\\
	& \overset{(d)} = \sum_{i=1}^n I(WY_{i+1}^nY_{1}^{i-1}X_iX_{1,i};Y_i) \notag\\
	&\qquad- \sum_{i=1}^nI(WY_{i+1}^{n}Y_{1}^{i-1};Y_{1,i}|X_iX_{1,i})+ n\epsilon_n \notag\\
	& \overset{(e)}= \sum_{i=1}^n I(T_iX_iX_{1,i};Y_i) - \sum_{i=1}^nI(T_i;Y_{1,i}|X_iX_{1,i})+ n\epsilon_n\notag\\
	& \overset{(f)}= \sum_{i=1}^n I(X_iX_{1,i};Y_i) - \sum_{i=1}^nI(T_i;Y_{1,i}|X_iX_{1,i}Y_i)+ n\epsilon_n, 
\end{align}
where $(a)$ follows from Fano's inequality~\cite{el2011network}, $(b)$ follows from the Csisz{\'a}r Sum Identity~\cite{el2011network}, $(c)$ follows since $X_i$ is a function of $W$ and $X_{1,i}$ is a function of $Y_{1}^{i-1}$, $(d)$ follows from that $I(WY_{1}^{i-1};Y_{1,i}|X_iX_{1,i})=0$ as $WY_{1}^{i-1}-X_iX_{1,i}-Y_{1,i}$ forms a Markov chain, $(e)$ follows from the definition of $T_i$, and $(f)$ follows from that $T_i-X_iX_{1,i}Y_{1,i}-Y_iS_iS_{d,i}$ forms a Markov chain. In addition, by the standard cut-set bound arguments~\cite{el2011network}, we have
\begin{align}
	nR & \le \sum_{i=1}^n I(X_i;Y_iY_{1,i}|X_{1,i})+ n\epsilon_n.
\end{align}

\par~Let $Q$ be a random variable independent of all other variables uniformly distributed over $[1:n]$ and set $X=X_Q$, $X_1=X_{1,Q}$, $S=S_{Q}$, $S_{d}=S_{d,Q}$, $Y=Y_Q$, $Y_1=Y_{1,Q}$, $T=(Q,T_Q)$. Since $Q-XX_1-YY_1$ forms a Markov chain, we have
\begin{align}
	&\sum_{i=1}^n I(X_iX_{1,i};Y_i) - \sum_{i=1}^nI(T_i;Y_{1,i}|X_iX_{1,i}Y_i) \notag\\
	&= n(I(X_QX_{1,Q};Y_Q|Q) - I(T_Q;Y_{1,Q}|X_QX_{1,Q}Y_QQ)) \notag\\
	&\le n\bigg(I(XX_{1};Y) - I(T;Y_1|XX_1Y)\bigg),
\end{align}
and 
\begin{align}
	\sum_{i=1}^n I(X_i;Y_iY_{1,i}|X_{1,i}) &= nI(X_Q;Y_QY_{1,Q}|X_{1,Q}Q) \notag\\
	&\le nI(X;YY_1|X_1).
\end{align}

\par~We proceed to proof the constraint on input distributions, i.e., $I(X_1;Y|X)\ge I(T;Y_1|XX_1Y)$. In~\eqref{equ:rateBoundKeyOne}, we have
\begin{align}\label{equ:ConsInputDisOne}
	\sum_{i=1}^n &I(WY_{i+1}^n;Y_i) \notag\\
	&=\sum_{i=1}^n I(X_iX_{1,i};Y_i) - \sum_{i=1}^nI(T_i;Y_{1,i}|X_iX_{1,i}Y_i).
\end{align}
We also note that
\begin{align}\label{equ:ConsInputDisTwo}
	\sum_{i=1}^n I(WY_{i+1}^n;Y_i) &\overset{(a)}= \sum_{i=1}^n I(WY_{i+1}^nX_i;Y_i)\notag\\
	& \ge \sum_{i=1}^n I(X_i;Y_i)
\end{align}
where $(a)$ follows since $X_i$ is a function of $W$. Thus, we have
\begin{align}
	\sum_{i=1}^n I(X_{1,i};Y_i|X_i) \ge \sum_{i=1}^nI(T_i;Y_{1,i}|X_iX_{1,i}Y_i),
\end{align}
i.e.,
\begin{align}
	I(X_{1,Q};Y_Q|X_QQ) \ge  I(T_Q;Y_{1,Q}|X_QX_{1,Q}Y_QQ).
\end{align}
Since $Q-XX_1-YY_1$ forms a Markov chain, we have
\begin{align}
	&I(X_{1,Q};Y_Q|X_QQ) \le I(X_1;Y|X),\\
	&I(T_Q;Y_{1,Q}|X_QX_{1,Q}Y_QQ) = I(T;Y_1|XX_1Y),
\end{align}
which indicates that there must be 
\begin{align}
	I(X_1;Y|X)\ge I(T;Y_1|XX_1Y).
\end{align}

\par~To bound the sensing performance, we use the following Lemma~\ref{lemma:dataProcessingEstimation}~\cite{choudhuri2013causal}.
\begin{lemma}\label{lemma:dataProcessingEstimation}
	(\textit{Data Processing Inequality for Parameter Estimation}~\cite{choudhuri2013causal}) Suppose $A\rightarrow B\rightarrow C$ form a Markov chain and $d(a,\hat{a})$ is a distortion measure. Then, for every reconstruction function $\hat{a}(b,c)$, there exist a reconstruction function $\hat{a}^*(b)$ such that 
	\begin{align}
		\mathbb{E}[d(A,\hat{a}^*(B))] \le \mathbb{E}[d(A,\hat{a}(B,C))].
	\end{align}
\end{lemma}

\par~Based on Lemma~\ref{lemma:dataProcessingEstimation}, we have
\begin{align}
	&\mathbb{E}[d(S_{d}^n,\hat{S}_{d}^n)] = \frac{1}{n}\sum_{i=1}^n\mathbb{E}[d(S_{{d},i},\hat{s}_{d,i}(Y^n))] \notag\\
	&\overset{(a)}\ge  \frac{1}{n}\sum_{i=1}^n \min_{\hat{s}^*_{d}(i,x_i,x_{1,i},y_{i},t_{i}) }\mathbb{E}[d(S_{d,i},\hat{s}_{d}^*(i,X_i,X_{1,i},Y_i,T_i))]\notag\\
	& = \min_{\hat{s}^*_{d}(x,x_{1},y,t) }\mathbb{E}[d(S_{d},\hat{s}_{d}^*(X,X_{1},Y,T))],
\end{align}
where $(a)$ follows from Lemma~\ref{lemma:dataProcessingEstimation} by identifying $S_{d,i}$ as $A$, $(X_i,X_{1,i},Y_i,T_i) = (X_i,X_{1,i},W,Y_{i}^n,Y_{1}^{i-1})$ as $B$, and $Y^{i-1}$ as $C$, and noting that $S_{d,i}-(X_i,X_{1,i},W,Y_{i}^n,Y_{1}^{i-1})-Y^{i-1}$ forms a Markov chain.
We proceed to show that an optimal estimator $\hat{s}_{d}^*(X,X_{1},Y,T)$ can be chosen as a deterministic function of $X,X_1,Y,T$ given in Theorem~\ref{theorem:outer}. We have
\begin{align}
	&\mathbb{E}_{XX_1YTS_{d}\hat{s}^*_{d}}[d(S_{d},\hat{s}^*_{d}(X,X_{1},Y,T))]\notag\\ &=\mathbb{E}_{XX_1YT}\bigg[\mathbb{E}_{S_{d}\hat{s}^*_{d}}[d(S_{d},\hat{s}^*_{d}(X,X_{1},Y,T))|XX_1YT]\bigg]\notag\\
	&\overset{(a)} = \sum_{x,x_1,y,t}P_{XX_1YT}(xx_1yt)\sum_{s_{d}}P_{S_{d}|XX_1YT}(s_{d}|xx_1yt)\notag\\
	&\qquad\qquad\quad \sum_{\hat{s}^*_{d}\in\hat{\mathcal{S}}_{d}}P_{\hat{S}^*_{d}|XX_1YT}(\hat{s}^*_{d}|xx_1yt)d(s_{d},\hat{s}^*_{d})\notag\\
	&\ge \sum_{x,x_1,y,t}P_{XX_1YT}(xx_1yt)\notag\\
	&\qquad\qquad\quad \min_{\hat{s}_{d}\in\hat{\mathcal{S}}_{d}}\sum_{s_{d}}P_{S_{d}|XX_1YT}(s_{d}|xx_1yt)d(s_{d},\hat{s}_{d})\notag\\
	& \overset{(b)}= \sum_{x,x_1,y,t}P_{XX_1YT}(xx_1yt)\notag\\
	&\qquad\qquad\quad\sum_{s_{d}}P_{S_{d}|XX_1YT}(s_{d}|xx_1yt)d(s_{d},\hat{S}_{d}(x,x_1,y,t))\notag\\
	&=\mathbb{E}_{XX_1YTS_{d}\hat{S}_{d}}[d(S_{d},\hat{S}_{d}(X,X_1,Y,T))],
\end{align}
where $(a)$ follows from the Markov chain $S_{d}-XX_1YT-\hat{s}_{d}(X,X_1,Y,T)$, $(b)$ follows by choosing the estimator~\eqref{equ:stateEstimator}.

Finally, we provide the proof for cardinality bound of auxiliary random variable $T$. Consider the following $|\mathcal{X}||\mathcal{X}_1|+1$ terms in Theorem~1:
	\begin{subequations}
		\begin{align}
			&P_{XX_1}(i),\  i = 1, \cdots,  |\mathcal{X}||\mathcal{X}_1|-1, \\
			&H(Y_1|XX_1YT),\\
			&\mathbb{E}[d(S_{d},\hat{S}_{d})]\le D.
		\end{align}
	\end{subequations}
	By the Support Lemma~[27], the alphabet size of $T$ can be upper-bounded by 
	\begin{align}
		|\mathcal{T}|\le |\mathcal{X}||\mathcal{X}_1|+1
	\end{align} 
	to preserve the above terms. In addition, we note that $P_{XX_1}(xx_1)$ determines 
	\begin{align}
		&P_{XX_1YY_1}(xx_1yy_1)\notag\\
		&\qquad=\sum_s P_{XX_1}(xx_1)P_S(s)P_{YY_1|XX_1S}(yy_1|xx_1s). 
	\end{align}
	When the joint distribution $P_{XX_1}$ is preserved, terms $I(X;YY_1|X_1)$, $I(XX_1;Y)$, $H(Y_1|X$ $X_1Y)$, and $I(X_1;Y|X)$ are also preserved. Thus, it suffices to consider $T$ whose alphabet $\mathcal{T}$ has cardinality $|\mathcal{T}|\le |\mathcal{X}||\mathcal{X}_1|+1$. 
	
\section{Proof of Theorem~\ref{theorem:inner}}\label{appendix:proofInner}
\par~We first present our achievable scheme for the case that $I(X_1;Y|X)>0$. Then, the discussion for the case that $I(X_1;Y|X)=0$ is provided. 

\subsection{Achievable Scheme for the Case that $I(X_1;Y|X)>0$}
\par~The transmission is performed in $B+1$ blocks, where each of the first $B$ blocks is of length $n$ channel uses, and the last block is of length $n_1=\frac{H(UAX_1Y_1)}{I(X_1;Y|X)}2n$ channel uses. A sequences of $B-1$ messages  $\{w_{(b)}\}_{b=1}^{B-1}$, each selected independently and uniformly distributed in $[1:2^{nR}]$, are transmitted over $B+1$ blocks. Each message $w_{(b)}$ is partitioned into two independent messages $\{w_{\text{c},(b)},w_{\text{p},(b)}\}\in[1:2^{nR_c}]\times[1:2^{nR_p}]$. The subscripts ``c'' and ``p'' here stand for ``common'' and ``private'', respectively. The relay in each block decodes the common message $w_{\text{c},(b-1)}$ sent by the source in the previous block~$b-1$. The source and the relay then cooperate in the current block~$b$ to collaboratively send $w_{\text{c},(b-1)}$. In addition, the relay also transmits a compressed version of its own side information, i.e., its own codeword, decoded common message, and received sequence, to the receiver, which can be leveraged for the message decoding and parameter estimation at the receiver. The destination decodes the message $\{w_{(b)}\}_{b=1}^{B-1}$ and outputs the estimated parameter sequence $\{\hat{S}_{d,i}\}_{i=1}^{nB}$ for the first $B$ blocks. 

\subsubsection{Codebook Generation}
Fix a joint distribution $P_UP_{A|U}$ $P_{X|UA}P_{X_1|U}P_{SS_{d}}P_{YY_1|XX_1S}
P_{V|UAX_1Y_1}P_{\hat{S}_{d}|XX_1YV}$,
where $P_{SS_{d}}P_{YY_1|XX_1S}$ is defined by the channel. Let $R=R_c+R_p$, $R_c\ge 0$, $R_p\ge 0$, and $R_v\ge 0$. For block~$b\in[1:B]$, the construction of codebook is as follows.
\begin{itemize}
	\item Generate $2^{nR_{c}}$ sequences $u_{(b)}^n(j_{b-1}), j_{b-1}\in[1:2^{nR_c}]$, i.i.d. according to $P_U(\cdot)$.
	\item For each index $j_{b-1}$, generate $2^{nR_c}$ sequences $a_{(b)}^n(j_{b-1},j_b)$, $j_{b}\in[1:2^{nR_c}]$, i.i.d. according to $P_{A|U}(\cdot|u)$.
	\item For each index pair $(j_{b-1},j_{b})$, generate $2^{nR_p}$ sequences $x_{(b)}^n(j_{b-1},j_b,k_b)$, $k_b\in[1:2^{nR_p}]$, i.i.d. according to $P_{X|UA}(\cdot|ua)$.
	\item For each index $j_{b-1}$, generate $2^{nR_v}$ sequences $x_{1,(b)}^n(j_{b-1},m_{b})$, $m_b\in[1:2^{nR_v}]$ i.i.d. according to $P_{X_1|U}(\cdot|u)$.
	\item For each index tuple $(j_{b-1},j_b,m_b)$, generate $2^{nR_{v}}$ sequences $v_{(b)}^n(j_{b-1},j_b,m_b,m_{b+1})$, $m_{b+1}\in[1:2^{nR_v}]$ i.i.d. according to $P_{V|UAX_1}(\cdot|uax_1)$.
\end{itemize}
The last block $b=B+1$ is devoted to the transmission of the compressed version of the side information at the relay of block~$B$, which contains no fresh message. The construction of codebook for block~$b=B+1$ is given as follows, where $P_{XX_1}$ is chosen by maximizing the $I(X_1;Y|X)$.
\begin{itemize}
	\item Generate one length$-n_1$ codeword $x_{(B+1)}^{n_1}(1)$, i.i.d. according to $P_X$.
	\item Generate $2^{nR_v}$ length$-n_1$ codewords $x_{1,(B+1)}^{n_1}(m_{B+1})$, $m_{B+1}\in[1:2^{nR_v}]$, i.i.d. according to $P_{X_1|X}$.
\end{itemize}
Reveal the codebooks to all parties.

\subsubsection{Encoding}
We set $j_0=1$, $m_1=1$, $j_B=1$, $k_B=1$. 
\begin{itemize}
	\item {\textbf{Block}}~$b=1$. The source and the relay send $x_{(1)}^n(1,j_1,k_1)$ and $x_{1,(1)}^n(1,1)$, respectively.
	\item {\textbf{Block}}~$b\in[2:B]$. The source sends $x_{(b)}^n(j_{b-1},j_b,k_b)$. At the end of block~$b-1$, the relay receives $y_{1,(b-1)}^n$ and finds a unique index $\hat{j}_{b-1}\in[1:2^{nR_c}]$ satisfying 
	\begin{align}\label{scheme:relayDecoding}
		\big( u_{(b-1)}^n(j_{b-2}), a_{(b-1)}^n(&j_{b-2},\hat{j}_{b-1}),x_{1,(b-1)}^n(j_{b-2},m_{b-1}),\notag\\
		&y_{1,(b-1)}^n
		\big)\in\mathcal{T}_{\epsilon}^n(UAX_1Y_1). 
	\end{align}
	If there is exactly one index $\hat{j}_{b-1}$ satisfying the above condition~\eqref{scheme:relayDecoding}, the relay sets $j_{b-1}=\hat{j}_{b-1}$. Otherwise, an error is declared. Once obtaining the correct index $j_{b-1}$, the relay finds an index $\hat{m}_{b}\in[1:2^{nR_v}]$ satisfying 
	\begin{align}
		\big( &u_{(b-1)}^n(j_{b-2}), a_{(b-1)}^n(j_{b-2},j_{b-1}), x_{1,(b-1)}^n(j_{b-2},m_{b-1}),\notag\\
		&\qquad y_{1,(b-1)}^n,v_{(b-1)}^n(j_{b-2},j_{b-1},m_{b-1},\hat{m}_{b})\big)\notag\\
		&\qquad\qquad\in\mathcal{T}_{\epsilon}^n(UAX_1Y_1V),
	\end{align}
	and sets $m_b=\hat{m}_b$. If there is no such pair, set $m_b=1$. Once obtaining the index $m_b$, the relay transmits $x_{1,(b)}^n(j_{b-1},m_{b})$ in block~$b$.
	\item {\textbf{Block}}~$b=B+1$. The source sends $x_{(B+1)}^{n_1}(1)$. For the relay, at the end of block~$B$, it receives $y_{1,(B)}^n$ and finds an index $\hat{m}_{B+1}\in[1:2^{nR_v}]$ satisfying 
	\begin{align}
		&\big( u_{(B)}^n(j_{B-1}), a_{(B)}^n(j_{B-1},1), x_{1,(B)}^n(j_{B-1},m_{B}), y_{1,(B)}^n,\notag\\
		&\quad v_{(B)}^n(j_{B-1},1,m_{B},\hat{m}_{B+1})\big)\in\mathcal{T}_{\epsilon}^n(UAX_1Y_1V),
	\end{align}
	and sets $m_{B+1}=\hat{m}_{B+1}$. If there is no such pair, set $m_{B+1}=1$. Once obtaining the index $m_{B+1}$, the relay transmits $x_{1,(B+1)}^{n_1}(m_{B+1})$ in block~$B+1$.
\end{itemize}

\subsubsection{Decoding}
Decoding begins at block~$B+1$ and proceeds backwards.
\begin{itemize}
	\item {\textbf{Block}}~$B+1$. The destination has $y_{(B+1)}^{n_1}$ and finds a unique index $\hat{m}_{B+1}\in[1:2^{nR_v}]$ such that 
	\begin{align}\label{scheme:receiverDecodingBlockB+1}
		\big( x_{(B+1)}^{n_1}(1),x_{1,{(B+1)}}^{n_1}(\hat{m}_{B+1}),y_{(B+1)}^{n_1}\big)\in\mathcal{T}_{\epsilon}^{n_1}(XX_1Y).
	\end{align}
	If there is exactly one index $\hat{m}_{B+1}$ satisfying~\eqref{scheme:receiverDecodingBlockB+1}, the destination sets $m_{B+1}=\hat{m}_{B+1}$. Otherwise, an error is declared.
	\item {\textbf{Block}}~$b\in[B:1]$. In block~$b$, the destination has at hand the pair  $(j_{b},m_{b+1})$ and $y_{(b)}^n$. It looks for a unique index tuple $(\hat{j}_{b-1},\hat{k}_{b},\hat{m}_{b})\in[1:2^{nR_c}]\times[1:2^{nR_p}]\times[1:2^{nR_v}]$ such that 
	\begin{align}\label{scheme:receiverDecodingBlockb}
		\big( &u_{(b)}^n(\hat{j}_{b-1}),a_{(b)}^n(\hat{j}_{b-1},j_{b}),x_{(b)}^n(\hat{j}_{b-1},j_{b},\hat{k}_{b}),\notag\\
		&\qquad x_{1,(b)}^n(\hat{j}_{b-1},\hat{m}_{b}),v_{(b)}^N(\hat{j}_{b-1},j_b,\hat{m}_b,m_{b+1}), y_{(b)}^n
		\big)\notag\\
		&\qquad\qquad\in\mathcal{T}_{\epsilon}^{n_1}(UAXX_1VY).
	\end{align}
	If there is exactly one index tuple $(\hat{j}_{b-1},\hat{k}_{b},\hat{m}_{b})$ satisfying~\eqref{scheme:receiverDecodingBlockb}, the destination sets $j_{b-1}=\hat{j}_{b-1}$, $k_{b}=\hat{k}_{b}$, and $m_{b} = \hat{m}_{b}$. Otherwise, an error is declared.
\end{itemize}
The message output at the destination is a sequence of pairs $(j_{b},k_{b})$,~$b\in[1:B-1]$.

\subsubsection{Parameter Estimation}
In block~$b\in[1:B]$, the destination has at hand the tuple $(j_{b-1},j_b,k_b,$ $m_b,m_{b+1})$ and $y_{(b)}^n$. It outputs 
\begin{align}
	\hat{s}_{d,(b)}^n = \hat{S}^{\otimes n}_d(x_{(b)}^n(&j_{b-1},j_b,k_b),x_{1,(b)}^n(j_{b-1},m_b),y_{(b)}^n,\notag\\
	&
	v_{(b)}^n(j_{b-1},j_b,m_b,m_{b+1})) 
\end{align}
by applying the symbol-by-symbol estimator defined in~\eqref{equ:stateEstimator}.

\subsubsection{The Analysis of Error Probability}
By standard information-theoretic arguments~\cite{el2011network} and letting $n,B\rightarrow\infty$, one can prove the error probability tends to zero if 
\begin{subequations}
	\begin{align}
		R_c &< I(A;Y_1|UX_1),  \\
		R_v &> I(V;Y_1|UAX_1), \\
		R_v &< 2H(UAX_1Y_1), \\
		R_c &+ R_p + R_v < I(XX_1;Y) + I(V;XY|UAX_1),  \\
		R_p &+ R_v < I(XX_1;Y|UA) + I(V;XY|UAX_1), \\
		R_p &< I(X;VY|UAX_1),  \\
		R_v &< I(X_1;Y|UAX) + I(V;XY|UAX_1). 
	\end{align}
\end{subequations}
Define $R = R_c + R_p$. By Fourier-Motzkin elimination~\cite{gattegno2016fourier}, one can eliminate auxiliary random variables $R_c$, $R_p$, $R_v$ and obtain the achievable region as stated in Theorem~\ref{theorem:inner}.

\subsubsection{The Analysis of Expected Distortion}
The total number of channel uses is $n_{\text{total}}=nB+n_1$, where $n_1=\frac{H(UAX_1Y_1)}{I(X_1;Y|X)}2n$ and $\frac{n_1}{n}$ is a finite number. Without loss in performance as $n,B\rightarrow\infty$, we focus on the average distortion in the first $B$ blocks. For any message $w=\{w_{(b)}\}_{b=1}^{B-1}$, the expected distortion in the first $B$ blocks is 
\begin{align}
	&\limsup_{n,B\rightarrow\infty}\Delta^{(nB)}(w)\overset{(a)}=\limsup_{n,B\rightarrow\infty}\mathbb{E}\bigg[\frac{1}{nB}\sum_{i=1}^{nB}d(S_{d,i},\hat{S}_{d,i})\bigg] \notag\\
	&\quad\overset{(b)}\le \limsup_{n,B\rightarrow\infty}\bigg( P_{\mathcal{E}}d_{\text{max}} + (1-P_{\mathcal{E}})(1+\epsilon)\mathbb{E}\bigg[d(S_{d},\hat{S}_{d})\bigg]\bigg)\notag\\
	&\quad\overset{(c)}\le\limsup_{n,B\rightarrow\infty}\bigg( P_{\mathcal{E}}d_{\text{max}} + (1-P_{\mathcal{E}})(1+\epsilon)\frac{D}{1+\epsilon}\bigg)\notag\\
	&\quad\overset{(d)}=D,
\end{align}
where $(a)$ follows from the definition in~\eqref{equ:defDistortionBlockN}, $(b)$ follows by applying the upper bound of the distortion function to the decoding error event and the typical average lemma~\cite{el2011network} to the successful decoding event,  $(c)$ follows from the random codebook generation and state estimating function that achieves $\frac{D}{1+\epsilon}$, and $(d)$ follows because $P_{\mathcal{E}}$ tends to zero as $n,B\rightarrow\infty$ if the constraints in Theorem~\ref{theorem:inner} hold.
Since the uniformly distributed messages are considered, we have
\begin{align}
	\limsup_{n,B\rightarrow\infty}\Delta^{(nB)} \le D,
\end{align} 
and thus as $n,B\rightarrow\infty$, we have
\begin{align}
	\limsup_{n,B\rightarrow\infty}\Delta^{(n_{\text{total}})} \le D.
\end{align}

\subsection{The Discussion for the Case that $I(X_1;Y|X)=0$}
We now turn to the case that $I(X_1;Y|X)=0$. Based on~\eqref{equ:ourRateConsThree} in Theorem~\ref{theorem:inner}, we have that the achievable rate $R$ in Theorem~\ref{theorem:inner} is upper bounded by
\begin{align}
	R \le I(X;Y) \le  I(X;Y x_1) = I(X;Y|x_1).
\end{align} 
We also have
\begin{align}\label{equ:Special}
	0 =I(X_1;Y|X) &\overset{(a)}\ge I(X_1;Y|UAX)\notag\\
	&\overset{(b)}\ge I(V;Y_1|UAXX_1Y)\notag\\
	&=H(Y_1|UAXX_1Y) - H(Y_1|UAXX_1YV)\notag\\
	&\overset{(c)}=H(Y_1|XX_1Y) - H(Y_1|UAXX_1YV)\notag\\
	&\overset{(d)}\ge H(Y_1|XX_1Y) - H(Y_1|XX_1YV)\notag\\
	& = I(V;Y_1|XX_1Y),
\end{align}
where $(a)$ follows from that conditioning reduces entropy and $UA-XX_1-Y$ forms a Markov chain,
$(b)$ follows from the inequality~\eqref{equ:ourConstraint} in Theorem~\ref{theorem:inner}, $(c)$ follows from that $UA-XX_1Y-Y_1$ forms a Markov chain, and $(d)$ follows from that conditioning reduces entropy. Thus, we have 
\begin{align}\label{eqU:specialCaseUselessV}
	0 &\overset{(a)}=I(V;S_d|XX_1YY_1) \notag\\
	&\overset{(b)}= I(V;S_dY_1|XX_1Y) \notag\\
	&\ge I(V;S_d|XX_1Y),
\end{align}
where $(a)$ follows from that $V-XX_1YY_1-S_d$ forms a Markov chain, and $(b)$ follows from~\eqref{equ:Special} that $I(V;Y_1|XX_1Y)=0$.
The results in~\eqref{eqU:specialCaseUselessV} reveals that when $I(X_1;Y|X)=0$, the state estimator in Theorem~\ref{theorem:inner} is equivalent to a deterministic function of $X,X_1,Y$.
Thus, the results in Theorem~\ref{theorem:inner} is upper bounded by
\begin{align}
	R \le \max_{x_1,P_{X|x_1}} I(X;Y|x_1), 
\end{align}
where $x_1,P_{X|x_1}$ are optimized over all $x_1\in\mathcal{X}_1$ and pmfs $P_{X|x_1}$ such that 
\begin{align}
	\mathbb{E}[d(S_d,\hat{S}_d)]\le D,
\end{align}
where $\hat{S}_d$ is given as 
\begin{align}
	\hat{S}_{d}(x,x_1,y) = \operatorname*{argmin}_{s'_d\in\mathcal{S}_d}\sum_{s_d\in\mathcal{S}_d}P_{S_d|XX_1Y}(s_d|x,x_1,y)d(s_d,s'_d).
\end{align}
This can be achieved by by letting the relay send a deterministic codeword $X_1=x_1$ and treat the relay channel as a point-to-point channel from source to the destination.

\section{Comparison between Our Scheme and the Chong-Motani-Garg Scheme\cite{chong2006generalized}}\label{appendix:discussionSimultaneouslyDecoding}
\par~In the Chong-Motani-Garg scheme~\cite{chong2006generalized}, the destination first jointly decodes the indices $(j_{b-1},m_b)$ of common message and compressed information, and then decodes the index $k_b$ of private message. By applying the Chong-Motani-Garg scheme~\cite{chong2006generalized} for bistatic ISAC over relay channels, one can obtain the following lower bound on $C(D)$:
\begin{theorem}\label{theorem:innerChong}
	The capacity-distortion function $C(D)$ is lower bounded by any positive rate $R$ satisfying
	\begin{subequations}
		\begin{align}
			R &\le I(A;Y_1|UX_1) + I(X;VY|UAX_1), \label{equ:rateConsInnerChongOne} \\
			R &\le I(XX_1;Y) - I(V;Y_1|UAXX_1Y), 
		\end{align}
	\end{subequations}
	and 
	\begin{align}\label{equ:chongConstraint}
		I(X_1;Y|UA) \ge I(V;Y_1|UAX_1Y), 
	\end{align}
	as well as the sensing distortion constraint
	\begin{align}
		\mathbb{E}[d(S_{d},\hat{S}_{d}(X,X_1,Y,V))]\le D, 
	\end{align}
	where the joint distribution of $UAXX_1SS_{d}YY_1V\hat{S}_{d}$ is 
	\begin{align}
		P_UP_{A|U}P_{X|UA}P_{X_1|U}P_{SS_{d}}&P_{YY_1|XX_1S}\notag\\
		&P_{V|UAX_1Y_1}P_{\hat{S}_{d}|XX_1YV}. 
	\end{align}
\end{theorem}

\par~For comparison, we first show that our achievable rate-distortion region in Theorem~\ref{theorem:inner} always includes that in Theorem~\ref{theorem:innerChong}.
\begin{theorem}\label{theorem:comparison}
	Let $I_{\text{R-D}}^{\text{our}}$ and $I_{\text{R-D}}^{\text{CMG}}$ denote the achievable regions in Theorem~\ref{theorem:inner} and Theorem~\ref{theorem:innerChong}, respectively. We have 
	\begin{align}
		I_{\text{R-D}}^{\text{CMG}} \subseteq I_{\text{R-D}}^{\text{our}}.
	\end{align}
	\begin{IEEEproof}
		By combining the results in $I_{\text{R-D}}^{\text{our}}$ and the constraint~\eqref{equ:chongConstraint}, one can obtain a new achievable region $I_{\text{R-D}}^{\text{tmp}}$ with
		\begin{align}
			I_{\text{R-D}}^{\text{tmp}} \subseteq I_{\text{R-D}}^{\text{our}}.
		\end{align}
		When the constraint~\eqref{equ:chongConstraint} holds, the constraint~\eqref{equ:ourConstraint} in Theorem~\ref{theorem:inner} is redundant since 
		\begin{align}
			I(X_1;Y|UAX) &= H(X_1|UAX) - H(X_1|UAXY)\notag\\
			&\overset{(a)}= H(X_1|UA) - H(X_1|UAXY)\notag\\
			&\overset{(b)}\ge H(X_1|UA) - H(X_1|UAY)\notag\\
			&= I(X_1;Y|UA),
		\end{align}
		and 
		\begin{align}
			I(V;Y_1|UA&X_1Y) = H(V|UAX_1Y) - H(V|UAX_1YY_1)\notag\\
			&\overset{(c)}\ge H(V|UAXX_1Y) - H(V|UAX_1YY_1)\notag\\
			&\overset{(d)}= H(V|UAXX_1Y) - H(V|UAXX_1YY_1)\notag\\
			&= I(V;Y_1|UAXX_1Y),
		\end{align}
		where $(a)$ follows from the Markov chain $X_1-UA-X$, $(b)$ and $(c)$ follow from that conditioning reduces entropy, and $(d)$ follows from the Markov chain $V-UAX_1Y_1-XY$. Moreover, one can also find that when the constraint~\eqref{equ:chongConstraint} holds, the rate constraint~\eqref{equ:ourRateConsTwo} in Theorem~\ref{theorem:inner} is redundant in the presence of constraint~\eqref{equ:ourRateConsOne} due to 
		\begin{align}
			&R\overset{(a)}\le I(A;Y_1|UX_1) + I(X;VY|UAX_1)   \notag\\
			&= I(A;Y_1|UX_1) + I(X;Y|UAX_1) + I(X;V|UAX_1Y) \notag\\
			&\qquad+ I(V;Y_1|UAX_1Y) -I(V;Y_1|UAX_1Y)\notag\\
			&\overset{(b)}= I(A;Y_1|UX_1) + I(X;Y|UAX_1) + I(X;V|UAX_1Y) \notag\\
			&\qquad+ I(X_1;Y|UA) -I(V;Y_1|UAX_1Y)\notag\\
			&= I(A;Y_1|UX_1) + I(XX_1;Y|UA) \notag\\
			&\qquad+ I(X;V|UAX_1Y) -I(V;Y_1|UAX_1Y)\notag\\
			&= I(A;Y_1|UX_1) + I(XX_1;Y|UA) \notag\\
			&\qquad- H(V|UAXX_1Y) + H(V|UAX_1YY_1)\notag\\
			&\overset{(c)}= I(A;Y_1|UX_1) + I(XX_1;Y|UA) \notag\\
			&\qquad- H(V|UAXX_1Y) + H(V|UAXX_1YY_1)\notag\\
			&= I(A;Y_1|UX_1) + I(XX_1;Y|UA) \notag\\
			&\qquad- I(V;Y_1|UAXX_1Y),
		\end{align}
		where $(a)$ follows from the rate constraint~\eqref{equ:ourRateConsOne}, $(b)$ follows from the constraint~\eqref{equ:chongConstraint}, and $(c)$ follows from the Markov chain $V-UAX_1Y_1-XY$.
		Thus, we have  
		\begin{align}
			I_{\text{R-D}}^{\text{CMG}} = I_{\text{R-D}}^{\text{tmp}},
		\end{align}
		which completes the proof.
	\end{IEEEproof}
\end{theorem}

\begin{remark}
	In the Chong-Motani-Garg scheme~\cite{chong2006generalized}, the destination only jointly decodes the indices of common message and compressed information, which results in a stricter constraint~\eqref{equ:chongConstraint} on achievable region and may restrict the performance of bistatic ISAC. 
\end{remark}

\par~We proceed to show that the inclusion in Theorem~\ref{theorem:comparison} can be strict , i.e., 
\begin{align}
	I_{\text{R-D}}^{\text{CMG}} \subsetneq I_{\text{R-D}}^{\text{our}}.
\end{align}
\begin{example}
	Consider a relay channel where the inputs $X,X_1$ are binary. The channel output at the relay is $Y_1=X_1\oplus S$, and the channel output at the destination is a binary tuple $Y=(X\oplus X_1,X\oplus N)$. The channel states $S,N$ are mutually independent binary random variables with $H(S)=1,H(N)=0.5$, i.e., $P_S\triangleq P_{S}(1)=0.5$ and $P_N\approx 0.11$. The sensing parameter is assumed as $S_d=S$, and the Hamming distortion $d(s_d,\hat{s}_d)=s_d\oplus\hat{s}_d$ is considered.
	
	In this example, the rate-distortion pair $(R,D)=(0.5,0)$ is in $I_{\text{R-D}}^{\text{our}}$, but not in $I_{\text{R-D}}^{\text{CMG}}$. In fact, if $D=0$ is in $I_{\text{R-D}}^{\text{CMG}}$, then $R$ must be zero.
	\begin{IEEEproof}
		We first show that $(R,D)=(0.5,0)$ is in $I_{\text{R-D}}^{\text{our}}$. For the results in Theorem~\ref{theorem:inner}, let $U=u^*$, $A=a^*$ almost surely for some specific values $u^*\in\mathcal{U}, a^*\in\mathcal{A}$, and $V=S$, $P_{X|UA=u^*a^*}=P_X=0.5$, $P_{X_1|U=u^*}=P_{X_1}=0.5$. One can find that $(R,D)=(0.5,0)$ is achievable.
		
		We proceed to show that in $I_{\text{R-D}}^{\text{CMG}}$, if $D=0$, i.e., $\hat{S}_{d}=S$, then $R$ must be zero. For the right hand side of~\eqref{equ:chongConstraint}, if $D=0$, we have
		\begin{align}
			I(V;Y_1|&UAX_1Y) = H(V|UAX_1Y) - H(V|UAX_1YY_1) \notag\\
			&\overset{(a)}=  H(V|UAX_1Y) - H(V|UAXX_1YY_1) \notag\\
			&\overset{(b)}\ge H(V|UAXX_1Y) - H(V|UAXX_1YY_1) \notag\\
			&\overset{(c)}= I(S;V|UAXX_1Y) \notag\\
			&\overset{(d)}= I(S;VUAXX_1Y) \notag\\
			&\overset{(e)}= I(S;VUAXX_1Y\hat{S}_{d})\notag\\
			& \ge I(S;\hat{S}_d) \notag\\
			&\overset{(f)}= 1,
		\end{align}
		where $(a)$ follows from the Markov chain $V-UAX_1Y_1-XY$, $(b)$ follows from that conditioning reduces entropy, $(c)$ follows from  that $Y_1=X_1\oplus S$ in the example, $(d)$ follows from that $I(S;UAXX_1Y)=0$ in the example, $(e)$ follows from that $\hat{S}_{d}$ is a deterministic function of $XX_1YV$, and $(f)$ follows from that $D=0$ and $H(S)=1$. We also notice that 
		\begin{align}
			I(V;Y_1|UAX_1Y) &=I (V;S|UAX_1Y) \le H(S)=1
		\end{align}
		since $Y_1=X_1\oplus S$ and $H(S)=1$. We thus have 
		\begin{align}
			I(V;Y_1|UAX_1Y) = 1.
		\end{align}
		We proceed to consider the left hand side of~\eqref{equ:chongConstraint} and have
		\begin{align}
			1&\overset{(a)}\le  I(X_1;Y|UA) \notag\\
			&= H(Y|UA)-H(Y|UAX_1)\notag\\
			&\overset{(b)}= H(X\oplus X_1,X\oplus N|UA) - H(X|UA) -H(N)\notag\\
			& = H(X\oplus X_1,X\oplus N,X_1|UA) \notag\\
			&\quad-H(X_1|UA,X\oplus X_1,X\oplus N) - H(X|UA) -H(N)\notag\\
			&= H(X_1|UA) +H(X|UA) +H(N) \notag\\
			&\quad-H(X_1|UA,X\oplus X_1,X\oplus N) - H(X|UA) -H(N)\notag\\
			&= H(X_1|UA) -H(X_1|UA,X\oplus X_1,X\oplus N)\notag\\
			& \le H(X_1)\notag\\
			& \le 1,
		\end{align}
		where $(a)$ follows from $I(V;Y_1|UAX_1Y) = 1$, $(b)$ follows from that $Y=(X\oplus X_1,X\oplus N)$ and $N$ is independent of $UAXX_1$. We thus have that the inequality~\eqref{equ:chongConstraint} holds when $D=0$ if and only if 
		\begin{align}
			& H(X_1|UA) = 1, \label{equ:XoneGIVENUAequalOne}\\
			& H(X_1|UA,X\oplus X_1,X\oplus N) = 0.
		\end{align}
		We then have
		\begin{align}\label{equ:X1OPLUSXgivenUAXOPLUSNequalONE}
			0 &= H(X_1|UA,X\oplus X_1,X\oplus N) \notag\\
			&= H(X_1,X_1\oplus X|UA,X\oplus N)\notag\\
			&\quad- H(X_1\oplus X|UA,X\oplus N)\notag\\
			&\overset{(a)}= H(X_1|UA) + H(X|UAX_1,X\oplus N) \notag\\
			&\quad- H(X_1\oplus X|UA,X\oplus N)\notag\\
			&\overset{(b)}= 1 + H(X,X\oplus N|UAX_1) \notag\\
			&\quad- H(X\oplus N|UAX_1) - H(X_1\oplus X|UA,X\oplus N)\notag\\
			&\overset{(c)}= 1 + H(X|UA) +H(N) \notag\\
			&\quad- H(X\oplus N|UA)- H(X_1\oplus X|UA,X\oplus N),
		\end{align}
		where $(a)$ follows from the Markov chain $X_1-UA-XN$, $(b)$ follows from~\eqref{equ:XoneGIVENUAequalOne}, $(c)$ follows from the Markov chain $X-UA-X_1$ and and $N$ is independent of $UAXX_1$. For the term $H(X_1\oplus X|UA,X\oplus N)$ in~\eqref{equ:X1OPLUSXgivenUAXOPLUSNequalONE}, we have
		\begin{align}
			H(X_1\oplus X|UA,X\oplus N) \le H(X_1\oplus X) \le  1
		\end{align}
		since $X,X_1$ are binary random variables, and 
		\begin{align}
			H(X_1\oplus X|UA,X\oplus N) &\overset{(a)}\ge H(X_1\oplus X|UAX,X\oplus N)\notag\\
			&=H(X_1|UAX,X\oplus N)\notag\\
			&\overset{(b)} = H(X_1|UA)\notag\\
			&\overset{(c)} = 1,
		\end{align}
		where $(a)$ follows from that conditioning reduces entropy, $(b)$ follows from the Markov chain $X_1-UA-XN$, and $(c)$ follows from~\eqref{equ:XoneGIVENUAequalOne}. We thus have that 
		\begin{align}
			H(X_1\oplus X|UA,X\oplus N) = 1
		\end{align}
		and combined with the results in~\eqref{equ:X1OPLUSXgivenUAXOPLUSNequalONE}, we have
		\begin{align}\label{equ:keyEquInExampleOneProof}
			H(X|UA) +H(N) - H(X\oplus N|UA) = 0. 
		\end{align}
		Given $H(N)=0.5$ and $X,X_1,N$ are binary random variables, we have that 
		\begin{align}
			H(X|UA)=0
		\end{align}
		by expanding the terms in~\eqref{equ:keyEquInExampleOneProof} for each $(u,a)\in\mathcal{U}\times\mathcal{A}$. Thus, based on the rate constraint~\eqref{equ:rateConsInnerChongOne} in Theorem~\ref{theorem:innerChong}, we have
		\begin{align}
			R &\le I(A;Y_1|UX_1)  + I(X;VY|UAX_1)\notag\\
			&\overset{(a)}= I(A;S|UX_1) + I(X;VY|UAX_1)\notag\\ 
			&\overset{(b)}= I(X;VY|UAX_1)\notag\\
			&\overset{(c)}\le H(X|UA)\notag\\
			&= 0,
		\end{align}
		where $(a)$ follows from that $Y_1=X_1\oplus S$, $(b)$ follows from that $S$ is independent of $UAX_1$, and $(c)$ follows from that conditioning reduces entropy and $H(X|UAX_1VY)\ge0$. We thus conclude that $R$ must be zero if $D=0$ is in $I_{\text{R-D}}^{\text{CMG}}$.
	\end{IEEEproof}
\end{example}

\section{Proof of the Relaxation for the Upper Bound}\label{appendix:relaxationOFouterBound}
We first show that the constraints in the relaxed upper bound must be satisfied when the results in Theorem~\ref{theorem:outer} hold, and then discuss the corresponding input distributions.

\par~First, given the results in Theorem~\ref{theorem:outer}, we have 
	\begin{align}
		R\le\min\{I(X;YY_1|X_1),I(XX_1;Y)\},
	\end{align}
	since the mutual information term $I(T;Y_1|XX_1Y)$ is nonnegative. Next, noting that the estimated state $\hat{S}_d$ is a function of $(X,X_1,Y,T)$ as stated in Theorem~\ref{theorem:outer}, we obtain
	\begin{align}
		I(S_d;\hat{S}_d) &\overset{(a)}\le I(S_d;XX_1YT) \notag\\
		&\le I(S_d;XX_1YY_1T)\notag\\
		& \overset{(b)}= I(S_d;XX_1YY_1)\notag\\
		&=  I(S_d;YY_1|XX_1),
	\end{align}
	where $(a)$ follows from the data-processing inequality~\cite{el2011network}, $(b)$ follows from the Markov chain $S_d-XX_1YY_1-T$, and
	\begin{align}
		I(&S_d;\hat{S}_d) \le I(S_d;XX_1YT)\notag\\
		&= I(S_d;YT|XX_1)\notag\\
		& = I(S_d;Y|XX_1) + I(S_d;T|XX_1Y)\notag\\
		& = I(S_d;Y|XX_1) + H(T|XX_1Y) - H(T|XX_1YS_d)\notag\\
		& \overset{(c)}\le I(S_d;Y|XX_1) + H(T|XX_1Y) - H(T|XX_1YY_1S_d)\notag\\
		& \overset{(d)}= I(S_d;Y|XX_1) + H(T|XX_1Y) - H(T|XX_1YY_1) \notag\\
		&= I(S_d;Y|XX_1) + I(T;Y_1|XX_1Y)\notag\\
		& \overset{(e)}\le I(S_d;Y|XX_1) + I(X_1;Y|X)\notag\\
		&= I(S_dX_1;Y|X),
	\end{align}
	where $(c)$ follows from the fact that conditioning reduces entropy,  $(d)$ follows from the Markov chain $S_d-XX_1YY_1-T$, and $(e)$ follows from the constraint $I(X_1;Y|X)\ge I(T;Y_1|XX_1Y)$ in Theorem~\ref{theorem:outer}. Hence, the constraints in relaxed upper bound are automatically satisfied when the results of Theorem~\ref{theorem:outer} hold.
	
\par~We now proceed to discuss the input distributions. To obtain a relaxed upper bound, we 
relax the conditional pmf $P_{\hat{S}_d|XX_1YT}$ as $P_{\hat{S}_d|XX_1YTS_d}$. Accordingly, the joint pmfs to be optimized becomes $P_{XX_1}P_{T|XX_1Y_1}P_{\hat{S}_d|XX_1YTS_d}$. Since all terms in the relaxed upper bound depend only on $P_{XX_1}$ or $P_{S_d\hat{S}_d}$, we can, without loss of generality, restrict the optimization to pmfs of the form $P_{XX_1}P_{\hat{S}_d|S_d}$.

\section{Proof of Example~\ref{example:optimalSensingGeneral}}\label{appendix:proofExampleoptimalSensingGeneral}
\par~We use $D$ to denote that expected distortion for sensing parameter $S_1$, i.e., 
\begin{align}
	D = \mathbb{E}[d(S_{\text{1}},\hat{S}_{\text{1}})] = \mathbb{E}[(S_{\text{1}}-\hat{S}_{\text{1}})^2].
\end{align}
Consider~\eqref{equ:descriptionCons} in Theorem~\ref{theorem:optimalSensing}, and we have
\begin{align}
	I(X_1;Y|X) &= h(Y|X) - h(Y|XX_1) \notag\\
	&\overset{(a)}= h(X_1+S_1+S_2|X) - h(S_1+S_2)\notag\\
	&\overset{(b)}\le h(X_1+S_1+S_2) - h(S_1+S_2)\notag\\
	&\overset{(c)}\le \frac{1}{2}\log(1+\frac{P_1}{\sigma^2_{s_1}+\sigma^2_{s_2}}),
\end{align}
and 
\begin{align}
	I(V;Y_1|XX_1Y) &\overset{(d)}=I(V\hat{S}_1;Y_1|XX_1Y) \notag\\
	&\ge I(\hat{S}_1;Y_1|XX_1Y)\notag\\
	& \overset{(e)}= I(\hat{S}_1;S_1|XX_1,S_1+S_2)\notag\\
	&\overset{(f)}\ge h(S_1|S_1+S_2) - h(S_1|\hat{S}_1)\notag\\
	&\overset{(g)}\ge h(S_1|S_1+S_2) - h(S_1-\hat{S}_1)\notag\\
	&\overset{(h)} \ge \frac{1}{2}\log(\frac{2\pi e\sigma^2_{s_1}\sigma^2_{s_2}}{\sigma^2_{s_1}+\sigma^2_{s_2}}) - \frac{1}{2}\log(2\pi e D) \notag\\
	&= \frac{1}{2}\log(\frac{\sigma^2_{s_1}\sigma^2_{s_1}}{(\sigma^2_{s_1}+\sigma^2_{s_2})D}),
\end{align}
where $(a)$ follows from that $Y=X+X_1+S_1+S_2$ and $S_1,S_2$ are independent of $X,X_1$ in Example~\ref{example:optimalSensingGeneral}, $(b)$ follows from that conditioning reduces entropy, $(c)$ follows from that the normal maximizes the entropy for a given variance, $(d)$ follows from the Markov chain $\hat{S}_1-XX_1YV-Y_1$, $(e)$ follows from $Y_1=X+S_1,Y=X+X_1+S_1+S_2$ in Example~\ref{example:optimalSensingGeneral}, $(f)$ follows from that $S_1S_2$ are independent of $XX_1$ and conditioning reduces entropy, $(g)$ follows from that conditioning reduces entropy, and $(h)$ follows from that the variance of $S_1-\hat{S}_1$ is no more than $D$ based on the quadratic distortion measure and the normal maximizes the entropy for a given variance. We thus have that 
\begin{align}
	D\ge \frac{\sigma^2_{s_1}\sigma^2_{s_2}}{P_1+\sigma^2_{s_1}+\sigma^2_{s_2}}.
\end{align}

\section{Proof of Proposition~\ref{proposition:relayChannelOptimalSensingDeterministic}}\label{appendix:proofForPropositionrelayChannelOptimalSensingDeterministic}
\par~Combining with the results in Theorem~\ref{theorem:optimalSensing} and the definition~\eqref{class:relayChannelOptimalSensingDeterministic} of channels $\mathcal{C}_1$, we have the Markov chain $S_{d}-XX_1Y-V$. The optimal estimated function~\eqref{equ:stateEstimator} degrades into 
\begin{align}
	\hat{S}_{d}(x,x_1,y) = \operatorname*{argmin}_{s'_{d}\in\hat{\mathcal{S}}_{d}}\sum_{s_{d}\in\mathcal{S}_{d}}P_{S_{d}|XX_1Y}(s_{d}|x,x_1,y)d(s_{d},s'_{d}),
\end{align}
and the minimum distortion in this case thus degrades into 
\begin{align}
	D_{\text{min}} = \min_{P_{XX_1}}\mathbb{E}[d(S_{d},\hat{S}_{d})],
\end{align}
since the choice of $V$ does not affect the $\mathbb{E}[d(S_{d},\hat{S}_{d})]$.
We further have
\begin{align}
	&\mathbb{E}_{XX_1YS_{d}\hat{S}^*_d}[d(S_{d},\hat{S}^*_{{d}}(X,X_{1},Y))]\notag\\
	&
	=\mathbb{E}_{XX_1Y}\bigg[\mathbb{E}_{S_{d}\hat{S}^*_d}[d(S_{d},\hat{S}^*_{{d}}(X,X_{1},Y))|XX_1Y]\bigg]\notag\\
	&= \sum_{x,x_1}P_{XX_1}(xx_1)\sum_{y}P_{Y|XX_1}(y|xx_1)\notag\\
	&\qquad\sum_{s_{d}}P_{S_{d}|XX_1Y}(s_{d}|xx_1y) d(s_{d},\hat{s}^*_{d}(x,x_1,y))\notag\\
	&\ge \min_{x,x_1}\sum_{y}P_{Y|XX_1}(y|xx_1)\notag\\
	&\qquad\sum_{s_{d}}P_{S_{d}|XX_1Y}(s_{d}|xx_1y) d(s_{d},\hat{s}^*_{d}(x,x_1,y))\notag\\
	&=\min_{x,x_1} \mathbb{E}[d(S_{d},\hat{S}_{d})].
\end{align}

\section{Proof of Proposition~\ref{proposition:OptimalSensingEstimateForward}}\label{appendix:ProofForPropositionOptimalSensingEstimateForward}
\par~The achievability follows by setting $V=\hat{S}_{d}$ for the results in Theorem~\ref{theorem:optimalSensing} and we have
\begin{align}
	I(X_1;Y|X) &\ge I(V;Y_1|XX_1Y) \notag\\
	&= H(Y_1|XX_1Y) - H(Y_1|XX_1VY) \notag\\
	&\overset{(a)}= H(Y_1|XX_1) - H(Y_1|XX_1V)\notag\\
	&= I(V;Y_1|XX_1)\notag\\
	&= I(\hat{S}_{d};Y_1|XX_1),
\end{align}
where $(a)$ follows from the Markov chains $Y-XX_1-Y_1$ and $Y-XX_1V-Y_1$.

\par~For the converse part, we first relax the constraint~\eqref{equ:descriptionCons} in Theorem~\ref{theorem:optimalSensing} as
\begin{align}\label{equ:relaxedConProposition2}
	I(X_1;Y|X)&\ge I(V;Y_1|XX_1Y) \notag\\
	&= H(Y_1|XX_1Y) - H(Y_1|XX_1VY)\notag\\
	&\overset{(a)}=H(Y_1|XX_1) - H(Y_1|XX_1VY)\notag\\
	&\overset{(b)}= H(Y_1|XX_1) - H(Y_1|XX_1VY\hat{S_{d}})\notag\\
	&\overset{(c)}\ge H(Y_1|XX_1) - H(Y_1|XX_1\hat{S_{d}})\notag\\
	&= I(\hat{S_{d}};Y_1|XX_1),
\end{align}
where $(a)$ follows from that Markov chain $Y-XX_1-Y_1$ holds for the channels in $\mathcal{C}_2$, $(b)$ follows from the fact that $\hat{S_{d}}$ is a deterministic function of $XX_1YV$, and $(c)$ follows from that conditioning reduces entropy. We proceed to relax the distribution of $\hat{S_{d}}$ as $P_{\hat{S_{d}}|XX_1Y_1YV}$ and obtain the following lower bound on $D_{\text{min}}$:
\begin{align}\label{optimization:DoneLower}
	D_{\text{min}} &\ge D^{1}_{\text{lower}} \notag\\
	&\triangleq  \min_{P_{XX_1}P_{V|XX_1Y_1}P_{\hat{S_{d}}|XX_1Y_1YV}}\mathbb{E}_{XX_1Y_1YVS_{d}\hat{S}_{d}}[d(S_{d},\hat{S}_{d})], 
\end{align}
where the minimum is over all pmfs $P_{XX_1}P_{V|XX_1Y_1}$ $P_{\hat{S_{d}}|XX_1Y_1YV}$ such that 
\begin{align}\label{cons:Donelower}
	I(X_1;Y|X)\ge I(\hat{S}_{d};Y_1|XX_1),
\end{align}
and $\hat{S}_{d}$ is not necessary to be a deterministic function of $X,X_1,Y_1,Y,V$. 
Next, we show that $D^{1}_{\text{lower}}$ is lower bounded by 
\begin{align}\label{optimization:DtwoLower}
	D^1_{\text{lower}} &\ge D^2_{\text{lower}} \notag\\
	&\triangleq  \min_{P_{XX_1}P_{V|XX_1Y_1}P_{\hat{S_{d}}|XX_1Y_1}}\mathbb{E}_{XX_1Y_1S_{d}\hat{S}_{d}}[d(S_{d},\hat{S}_{d})], 
\end{align}
where the minimum is over all pmfs $P_{XX_1}P_{V|XX_1Y_1}$ $P_{\hat{S_{d}}|XX_1Y_1}$ such that 
\begin{align}\label{cons:DtwoLower}
	I(X_1;Y|X)\ge I(\hat{S}_{d};Y_1|XX_1),
\end{align}
and $\hat{S}_{d}$ is not necessary to be a deterministic function of $X,X_1,Y_1$. 
Our key idea is to show that we can find a feasible solution for optimization~\eqref{optimization:DtwoLower} achieves $D^1_{\text{lower}}$.
Let $P^1_{XX_1}P^1_{V|XX_1Y_1}P^1_{\hat{S}_{d}|XX_1Y_1YV}$ denote the optimum distribution that achieves $D^1_{\text{lower}}$ in~\eqref{optimization:DoneLower}. We have
\begin{align}\label{equ:introduceNewDistribution}
	&P^1_{\hat{S}_{d}|XX_1Y_1}(\hat{s}_{d}|xx_1y_1) = 
	\sum_{y,v}P^1_{\hat{S}_{d}YV|XX_1Y_1}(\hat{s}_{d}yv|xx_1y_1)\notag\\
	&= 
	\sum_{y,v}P^1_{V|XX_1Y_1}(v|xx_1y_1)P_{Y|XX_1Y_1V}(y|xx_1y_1v)\notag\\
	&\qquad \qquad P^1_{\hat{S}_{d}|XX_1Y_1YV}(\hat{s}_{d}|xx_1y_1yv)\notag\\
	&\overset{(a)}= \sum_{y,v}P^1_{V|XX_1Y_1}(v|xx_1y_1)P_{Y|XX_1}(y|xx_1)\notag\\
	&\qquad\qquad P^1_{\hat{S}_{d}|XX_1Y_1YV}(\hat{s}_{d}|xx_1y_1yv),
\end{align}
where $(a)$ follows from that Markov chain $Y-XX_1-Y_1V$ holds for the channels in $\mathcal{C}_2$, and $P_{Y|XX_1}(y|xx_1)$ is defined by the channel since 
\begin{align}
	P_{Y|XX_1}(y|xx_1) = \sum_{s_2}P_{S_2}(s_2)P_{Y|XX_1S_2}(y|xx_1s_2).
\end{align}
The terms in constraints~\eqref{cons:Donelower} and~\eqref{cons:DtwoLower} are only dependent on the pmf $P_{XX_1}P_{\hat{S}_{d}|XX_1Y_1}$. Since $P^1_{XX_1}P^1_{V|XX_1Y_1}$ $P^1_{\hat{S}_{d}|XX_1Y_1YV}$ is a feasible distribution for the optimization problem~\eqref{optimization:DoneLower}, we have that $P^1_{XX_1}P^1_{V|XX_1Y_1}P^1_{\hat{S}_{d}|XX_1Y_1}$ is a feasible distribution for the optimization problem~\eqref{optimization:DtwoLower}. We also have
\begin{align}
	&\mathbb{E}_{XX_1Y_1S_{d}\hat{S}_{d}}[d(S_{d},\hat{S}_{d})]\bigg|_{P^1_{XX_1}P^1_{V|XX_1Y_1}P^1_{\hat{S}_{d}|XX_1Y_1}}\notag\\
	&\overset{(a)}= \sum_{x,x_1,y_1,s_{d},\hat{s}_{d}}P_{XX_1Y_1S_{d}}(xx_1y_1s_{d})\notag\\
	&\qquad\quad P_{\hat{S}_{d}|XX_1Y_1}(\hat{s}_{d}|xx_1y_1)d(s_{d},\hat{s}_{d})\bigg|_{P^1_{XX_1}P^1_{V|XX_1Y_1}P^1_{\hat{S}_{d}|XX_1Y_1}}\notag\\
	&\overset{(b)}= \sum_{x,x_1,y_1,s_{d},\hat{s}_{d}} P^1_{XX_1}(xx_1)P_{Y_1|XX_1}(y_1|xx_1)\notag\\
	&\qquad\quad P_{S_{d}|XX_1Y_1}(s_{d}|xx_1y_1) \sum_{y,v}P^1_{V|XX_1Y_1}(v|xx_1y_1)\notag\\
	&\qquad\qquad P_{Y|XX_1}(y|xx_1)P^1_{\hat{S}_{d}|XX_1Y_1YV}(\hat{s}_{d}|xx_1y_1yv)d(s_{d},\hat{s}_{d})\notag\\
	&\overset{(c)}= \sum_{x,x_1,y_1,y,v,s_{d},\hat{s}_{d}}P^1_{XX_1}(xx_1)P_{Y_1Y|XX_1}(y_1y|xx_1)\notag\\
	&\qquad\quad P^1_{V|XX_1Y_1}(v|xx_1y_1) P_{S_{d}|XX_1Y_1}(s_{d}|xx_1y_1)\notag\\
	&\qquad\qquad P^1_{\hat{S_{d}}|XX_1Y_1YV}(\hat{s}_{d}|xx_1y_1yv)d(s_{d},\hat{s}_{d})\notag\\
	&\overset{(d)}= \sum_{x,x_1,y_1,y,v,s_{d},\hat{s}_{d}}P^1_{XX_1}(xx_1)P_{Y_1Y|XX_1}(y_1y|xx_1)\notag\\
	&\qquad\quad P^1_{V|XX_1Y_1}(v|xx_1y_1) P_{S_{d}|XX_1Y_1YV}(s_{d}|xx_1y_1yv)\notag\\
	&\qquad\qquad P^1_{\hat{S}_{d}|XX_1Y_1YV}(\hat{s}_{d}|xx_1y_1yv)d(s_{d},\hat{s}_{d})\notag\\
	&= \mathbb{E}_{XX_1Y_1YVS_{d}\hat{S}_{d}}[d(S_{d},\hat{S}_{d})]\bigg|_{P^1_{XX_1}P^1_{V|XX_1Y_1}P^1_{\hat{S}_{d}|XX_1Y_1YV}}\notag\\
	&= D^1_{\text{lower}},
\end{align}
where $(a)$ follows from the Markov chain $\hat{S}_{d}-XX_1Y_1-S_{d}$ in optimization~\eqref{optimization:DtwoLower}, $(b)$ follows from~\eqref{equ:introduceNewDistribution}, $(c)$ follows from the Markov chain $Y-XX_1-Y_1$ holds for the channels in $\mathcal{C}_2$, and $(d)$ follows from the Markov chain $S_{d}-XX_1Y_1-YV$ holds for the channels in $\mathcal{C}_2$. Thus, we have
\begin{align}
	D_{\text{min}}\ge D^1_{\text{lower}} \ge D^2_{\text{lower}},
\end{align}
since there already exists a feasible distribution $P^1_{XX_1}$ $P^1_{V|XX_1Y_1}P^1_{\hat{S}_{d}|XX_1Y_1}$ achieves $D^1_{\text{lower}}$ for the optimization problem~\eqref{optimization:DtwoLower}. We also notice that for the optimization~\eqref{optimization:DtwoLower}, all terms are independent of $P_{V|XX_1Y_1}$. Without loss of generality, we can restrict the
optimization over the choice of pmfs of the form $P_{XX_1}P_{\hat{S}_{d}|XX_1Y_1}$, i.e., 
\begin{align}
	D_{\text{min}} &\ge D^2_{\text{lower}} \notag\\
	&\triangleq  \min_{P_{XX_1}P_{\hat{S}_{d}|XX_1Y_1}}\mathbb{E}_{XX_1Y_1S_{d}\hat{S}_{d}}[d(S_{d},\hat{S}_{d})], 
\end{align}
where the minimum is over all pmfs $P_{XX_1}P_{\hat{S_{d}}|XX_1Y_1}$ such that 
\begin{align}
	I(X_1;Y|X)\ge I(\hat{S}_{d};Y_1|XX_1).
\end{align}

\section{Proof of Example~\ref{example:optimalSensingEstimatingForward}}\label{appendix:proofOFExample3}
\par~We first use $D$ to denote that expected distortion for sensing parameter $S_1$, i.e., 
\begin{align}
	D = \mathbb{E}[d(S_{\text{1}},\hat{S}_{\text{1}})] = \mathbb{E}[(S_{\text{1}}-\hat{S}_{\text{1}})^2].
\end{align}
Consider~\eqref{equ:consProposition2} in Proposition~\ref{proposition:OptimalSensingEstimateForward}, and we have
\begin{align}
	I(X_1;Y|X) &= h(Y|X) - h(Y|XX_1) \notag\\
	&\overset{(a)}= h(X_1+S_2|X) - h(S_2)\notag\\
	&\overset{(b)}\le h(X_1+S_2) - h(S_2)\notag\\
	&\overset{(c)}\le \frac{1}{2}\log(1+\frac{P_1}{\sigma^2_{s_2}}),
\end{align}
and 
\begin{align}
	I(\hat{S}_{1};Y_1|XX_1) &\overset{(d)}=I(\hat{S}_1;S_1|XX_1) \notag\\
	&= h(S_1|XX_1) - h(S_1|XX_1\hat{S}_1)\notag\\
	&\overset{(e)}= h(S_1) - h(S_1|XX_1\hat{S}_1)\notag\\
	&= h(S_1) - h(S_1-\hat{S}_1|XX_1\hat{S}_1)\notag\\
	&\overset{(f)}\ge h(S_1) - h(S_1-\hat{S}_1)\notag\\
	&\overset{(g)}\ge \frac{1}{2}\log(\frac{\sigma^2_{s_1}}{D}),
\end{align}
where $(a)$ follows from that $Y=X+X_1+S_2$ in Example~\ref{example:optimalSensingEstimatingForward}, $(b)$ follows from that conditioning reduces entropy, $(c)$ follows from that the normal maximizes the entropy for a given variance, $(d)$ follows from that $Y_1=X+S_1$ in Example~\ref{example:optimalSensingEstimatingForward}, $(e)$ follows from that $S_1$ is independent of $X,X_1$, $(f)$ follows from that conditioning reduces entropy, and $(g)$ follows from that the normal maximizes the entropy for a given variance and the variance of $S_1-\hat{S}_1$ is no more than $D$ based on the quadratic distortion measure. We thus have
\begin{align}
	D\ge \frac{\sigma^2_{s_1}\sigma^2_{s_2}}{P_1+\sigma^2_{s_2}}.
\end{align}

\section{Proof of Theorem~\ref{theorem:cover-kimRelayChannel}}\label{appendix:cover-kimRelayChannel}
\par~We first show that the results in Theorem~\ref{theorem:cover-kimRelayChannel} are achievable. Setting $U=u^*$, $A=a^*$ almost surely some specific values $u^*\in\mathcal{U}$, $a^*\in\mathcal{A}$ and $V=Y_1$. The joint pmf of random variables $XX_1SS_dY_rY_dY_1\hat{S}_d$ is thus 
\begin{align}
	P_{X}P_{X_1}P_{SS_d}P_{Y_dY_1|XS}P_{Y_r|X_1}P_{\hat{S}_d|XY_d},
\end{align}
and we have
\begin{align}
	I(A;&Y_1|UX_1) + I(X;VY|UAX_1) = I(X;Y_1Y|X_1)\notag\\
	&\overset{(a)}= I(X;Y_1Y_rY_d|X_1)\notag\\
	&= I(X;Y_1Y_d|X_1) + I(X;Y_r|X_1Y_1Y_d)\notag\\
	&\overset{(b)}=I(X;Y_1Y_d|X_1) \notag\\
	&= H(Y_1Y_d|X_1) - H(Y_1Y_d|X_1X)\notag\\
	&\overset{(c)}= H(Y_1Y_d) - H(Y_1Y_d|X)\notag\\
	&= I(X;Y_1Y_d),
\end{align}
and
\begin{align}
	I(A;&Y_1|UX_1) + I(XX_1;Y|UA) - I(V;Y_1|UAXX_1Y) \notag\\
	&= I(XX_1;Y)-I(Y_1;Y_1|XX_1Y)\notag\\
	&\overset{(d)}= I(XX_1;Y_rY_d)-I(Y_1;Y_1|XX_1Y_rY_d)\notag\\
	&\overset{(e)}= I(XX_1;Y_rY_d)\notag\\
	&= I(XX_1;Y_r) + I(XX_1;Y_d|Y_r)\notag\\
	&= H(Y_r)-H(Y_r|XX_1) +H(Y_d|Y_r) - H(Y_d|Y_rXX_1) \notag\\
	&\overset{(f)}=H(Y_r)-H(Y_r|X_1) +H(Y_d) - H(Y_d|X) \notag\\
	&= I(X_1;Y_r) + I(X;Y_d),
\end{align}
where $(a)$ follows from that $Y=(Y_r,Y_d)$ for the channels in $\mathcal{C}_4$, $(b)$ follows from the Markov chain $Y_r-X_1Y_1Y_d-X$, $(c)$ follows from that $Y_1Y_d$ are independent of $X_1$ and the Markov chain $Y_1Y_d-X-X_1$, $(d)$ follows from that $Y=(Y_r,Y_d)$ for the channels in $\mathcal{C}_4$, $(e)$ follows from that $Y_1=f(X,Y_d)$ is a deterministic function of $X,Y_d$, $(f)$ follows from the Markov chains $Y_r-X_1-X$, $Y_d-X-Y_rX_1$ and $Y_d,Y_r$ are mutually independent due to the joint pmf $P_{X}P_{X_1}P_{SS_d}P_{Y_dY_1|XS}P_{Y_r|X_1}P_{\hat{S}_d|XY_d}$. Note that with the above choice of random variables $U,A,V$, the constraint~\eqref{equ:ourConstraint} in Theorem~\ref{theorem:inner} is inactive, and the state estimator degrades into 
\begin{align}
	&\hat{S}_{d}(x,x_1,y,v) \notag\\
	&= \operatorname*{argmin}_{s'_{d}\in\hat{\mathcal{S}}_{d}}\sum_{s_{d}\in\mathcal{S}_{d}}P_{S_{d}|XX_1YV}(s_{d}|x,x_1,y,v)d(s_{d},s'_{d})\notag\\
	&\overset{(a)}=\operatorname*{argmin}_{s'_{d}\in\hat{\mathcal{S}}_{d}}\sum_{s_{d}\in\mathcal{S}_{d}}P_{S_{d}|XX_1Y_rY_dY_1}(s_{d}|x,x_1,y_r,y_d,y_1)d(s_{d},s'_{d})\notag\\
	&\overset{(b)}= \operatorname*{argmin}_{s'_{d}\in\hat{\mathcal{S}}_{d}}\sum_{s_{d}\in\mathcal{S}_{d}}P_{S_{d}|XY_d}(s_{d}|x,y_d)d(s_{d},s'_{d}),
\end{align}
where $(a)$ follows from that $Y=(Y_r,Y_d)$ for the channels in $\mathcal{C}_4$, $(b)$ follows from the Markov chain $S_{d}-XY_d-X_1Y_1Y_r$ for the channels in ${\mathcal C}_4$.

\par~We proceed to consider the converse part. Note that for the channels in ${\mathcal C}_4$, we have the Markov chain $S_{d}-XY_d-X_1Y_rY_1T$. Thus, the state estimator ~\eqref{equ:stateEstimator} degrades into a deterministic function of $X,Y_d$. Next, we relax the rate constraints in Theorem~\ref{theorem:outer} as 
\begin{align}
	I(X;YY_1|X_1) & \overset{(a)}= I(X;Y_rY_dY_1|X_1) \notag\\
	&\overset{(b)}=I(X;Y_1Y_d|X_1) \notag\\
	&= H(Y_1Y_d|X_1) - H(Y_1Y_d|XX_1)\notag\\
	& \overset{(c)}\le H(Y_1Y_d) - H(Y_1Y_d|X)\notag\\
	&= I(X;Y_1Y_d),
\end{align}
and 
\begin{align}
	I(&XX_1;Y)-I(T;Y_1|XX_1Y) \notag\\
	&\le I(XX_1;Y) \notag\\
	&\overset{(d)}= I(XX_1;Y_rY_d) \notag\\
	&= I(XX_1;Y_r) + I(XX_1;Y_d|Y_r)\notag\\
	&= H(Y_r)-H(Y_r|XX_1) +H(Y_d|Y_r) - H(Y_d|Y_rXX_1) \notag\\
	&\overset{(e)}\le H(Y_r)-H(Y_r|X_1) +H(Y_d) - H(Y_d|X) \notag\\
	&= I(X_1;Y_r) + I(X;Y_d),
\end{align}
where $(a)$ follows from that $Y=(Y_r,Y_d)$ for the channels in $\mathcal{C}_4$, $(b)$ follows from the Markov chain $Y_r-X_1Y_1Y_d-X$, $(c)$ follows from that conditioning reduces entropy and the Markov chain $Y_1Y_d-X-X_1$, $(d)$ follows from that $Y=(Y_r,Y_d)$ for the channels in $\mathcal{C}_4$, $(e)$ follows from the Markov chains $Y_r-X_1-X$, $Y_d-X-X_1Y_r$ and conditioning reduces entropy. We proceed to ignore the constraint
$I(X_1;Y|X)\ge I(T;Y_1|XX_1Y)$
in Theorem~\ref{theorem:outer} and obtain all constraints shown in Theorem~\ref{theorem:cover-kimRelayChannel}. Note that all terms depend on $P_X$ or $P_{X_1}$. Without loss of generality, we can restrict the optimization over the choice of joint pmf of the form $P_{X}P_{X_1}$.

\bibliographystyle{IEEEtran}
\bibliography{reference}

\begin{thebibliography}{10}
\providecommand{\url}[1]{#1}
\csname url@samestyle\endcsname
\providecommand{\newblock}{\relax}
\providecommand{\bibinfo}[2]{#2}
\providecommand{\BIBentrySTDinterwordspacing}{\spaceskip=0pt\relax}
\providecommand{\BIBentryALTinterwordstretchfactor}{4}
\providecommand{\BIBentryALTinterwordspacing}{\spaceskip=\fontdimen2\font plus
\BIBentryALTinterwordstretchfactor\fontdimen3\font minus
  \fontdimen4\font\relax}
\providecommand{\BIBforeignlanguage}[2]{{%
\expandafter\ifx\csname l@#1\endcsname\relax
\typeout{** WARNING: IEEEtran.bst: No hyphenation pattern has been}%
\typeout{** loaded for the language `#1'. Using the pattern for}%
\typeout{** the default language instead.}%
\else
\language=\csname l@#1\endcsname
\fi
#2}}
\providecommand{\BIBdecl}{\relax}
\BIBdecl

\bibitem{liu2022survey}
A.~Liu, Z.~Huang, M.~Li, Y.~Wan, W.~Li, T.~X. Han, C.~Liu, R.~Du, D.~K.~P. Tan,
  J.~Lu \emph{et~al.}, ``A survey on fundamental limits of integrated sensing
  and communication,'' \emph{IEEE Communications Surveys \& Tutorials},
  vol.~24, no.~2, pp. 994--1034, 2022.

\bibitem{liu2022integrated}
F.~Liu, Y.~Cui, C.~Masouros, J.~Xu, T.~X. Han, Y.~C. Eldar, and S.~Buzzi,
  ``Integrated sensing and communications: Toward dual-functional wireless
  networks for 6g and beyond,'' \emph{IEEE {J}ournal on {S}elected {A}reas in
  {C}ommunications}, vol.~40, no.~6, pp. 1728--1767, 2022.

\bibitem{kobayashi2018joint}
M.~Kobayashi, G.~Caire, and G.~Kramer, ``Joint state sensing and communication:
  Optimal tradeoff for a memoryless case,'' in \emph{Proc. IEEE ISIT}, 2018,
  pp. 111--115.

\bibitem{liu2022information}
Y.~Liu, M.~Li, A.~Liu, J.~Lu, and T.~X. Han, ``Information-theoretic limits of
  integrated sensing and communication with correlated sensing and channel
  states for vehicular networks,'' \emph{IEEE Transactions on Vehicular
  Technology}, vol.~71, no.~9, pp. 10\,161--10\,166, 2022.

\bibitem{kobayashi2019joint}
M.~Kobayashi, H.~Hamad, G.~Kramer, and G.~Caire, ``Joint state sensing and
  communication over memoryless multiple access channels,'' in \emph{Proc. IEEE
  ISIT}, 2019, pp. 270--274.

\bibitem{ahmadipour2023information}
M.~Ahmadipour and M.~Wigger, ``An information-theoretic approach to
  collaborative integrated sensing and communication for two-transmitter
  systems,'' \emph{IEEE Journal on Selected Areas in Information Theory},
  vol.~4, pp. 112--127, 2023.

\bibitem{liu2023Globecom}
Y.~Liu, M.~Li, A.~Liu, and L.~Ong, ``Improved information-theoretic bound for
  multiple-access integrated sensing and communication systems,'' in
  \emph{Proc. IEEE GLOBECOM}, 2023, pp. 7369--7374.

\bibitem{liu2022generalized}
Y.~Liu, M.~Li, A.~Liu, L.~Ong, and A.~Yener, ``Fundamental limits of
  multiple-access integrated sensing and communication systems,'' \emph{IEEE
  Transactions on Information Theory}, vol.~71, no.~6, pp. 4317--4341, 2025.

\bibitem{ahmadipour2022information}
M.~Ahmadipour, M.~Kobayashi, M.~Wigger, and G.~Caire, ``An
  information-theoretic approach to joint sensing and communication,''
  \emph{IEEE Transactions on Information Theory}, vol.~70, no.~2, pp.
  1124--1146, 2022.

\bibitem{liu2024ICC}
Y.~Liu, M.~Li, Y.~Han, and L.~Ong, ``Fundamental limits of integrated sensing
  and communication over interference channels,'' \emph{IEEE Journal on
  Selected Areas in Information Theory}, vol.~7, pp. 1--15, 2026.

\bibitem{xiong2023fundamental}
Y.~Xiong, F.~Liu, Y.~Cui, W.~Yuan, T.~X. Han, and G.~Caire, ``On the
  fundamental tradeoff of integrated sensing and communications under
  {G}aussian channels,'' \emph{IEEE Transactions on Information Theory},
  vol.~69, no.~9, pp. 5723--5751, 2023.

\bibitem{gunlu2023secure}
O.~G{\"u}nl{\"u}, M.~R. Bloch, R.~F. Schaefer, and A.~Yener, ``Secure
  integrated sensing and communication,'' \emph{IEEE Journal on Selected Areas
  in Information Theory}, vol.~4, pp. 40--53, 2023.

\bibitem{welling2024transmitter}
T.~Welling, O.~G{\"u}nl{\"u}, and A.~Yener, ``Transmitter actions for secure
  integrated sensing and communication,'' in \emph{Proc. IEEE ISIT}, 2024, pp.
  2580--2585.

\bibitem{nikbakht2024integrated}
H.~Nikbakht, M.~Wigger, S.~Shamai, and H.~V. Poor, ``Integrated sensing and
  communication in the finite blocklength regime,'' in \emph{Proc. IEEE ISIT},
  2024, pp. 2790--2795.

\bibitem{zhang2011joint}
W.~Zhang, S.~Vedantam, and U.~Mitra, ``Joint transmission and state estimation:
  A constrained channel coding approach,'' \emph{IEEE Transactions on
  Information Theory}, vol.~57, no.~10, pp. 7084--7095, 2011.

\bibitem{salimi2017capacity}
A.~Salimi, W.~Zhang, S.~Vedantam, and U.~Mitra, ``The capacity-distortion
  function for multihop channels with state,'' in \emph{Proc. IEEE ISIT}, 2017,
  pp. 2228--2232.

\bibitem{jiao2024rate}
T.~Jiao, K.~Wan, Z.~Wei, Y.~Geng, Y.~Li, Z.~Yang, and G.~Caire,
  ``Information-theoretic limits of bistatic integrated sensing and
  communication,'' \emph{IEEE Transactions on Information Theory}, vol.~71,
  no.~12, pp. 9302--9318, 2025.

\bibitem{ahmadipour2023strong}
M.~Ahmadipour, M.~Wigger, and S.~Shamai, ``Strong converses for memoryless
  bi-static {ISAC},'' in \emph{Proc. IEEE ISIT}, 2023, pp. 1818--1823.

\bibitem{sutivong2005channel}
A.~Sutivong, M.~Chiang, T.~M. Cover, and Y.-H. Kim, ``Channel capacity and
  state estimation for state-dependent gaussian channels,'' \emph{IEEE
  Transactions on Information Theory}, vol.~51, no.~4, pp. 1486--1495, 2005.

\bibitem{kim2008state}
Y.-H. Kim, A.~Sutivong, and T.~M. Cover, ``State amplification,'' \emph{IEEE
  Transactions on Information Theory}, vol.~54, no.~5, pp. 1850--1859, 2008.

\bibitem{choudhuri2013causal}
C.~Choudhuri, Y.-H. Kim, and U.~Mitra, ``Causal state communication,''
  \emph{IEEE Transactions on Information Theory}, vol.~59, no.~6, pp.
  3709--3719, 2013.

\bibitem{tian2015gaussian}
C.~Tian, B.~Bandemer, and S.~S. Shitz, ``Gaussian state amplification with
  noisy observations,'' \emph{IEEE Transactions on Information Theory},
  vol.~61, no.~9, pp. 4587--4597, 2015.

\bibitem{ramachandran2019joint}
V.~Ramachandran, S.~R.~B. Pillai, and V.~M. Prabhakaran, ``Joint state
  estimation and communication over a state-dependent gaussian multiple access
  channel,'' \emph{IEEE Transactions on Communications}, vol.~67, no.~10, pp.
  6743--6752, 2019.

\bibitem{cover1979capacity}
T.~Cover and A.~E. Gamal, ``Capacity theorems for the relay channel,''
  \emph{IEEE Transactions on Information Theory}, vol.~25, no.~5, pp. 572--584,
  1979.

\bibitem{chong2006generalized}
H.-F. Chong, M.~Motani, and H.~K. Garg, ``Generalized backward decoding
  strategies for the relay channel,'' \emph{IEEE Transactions on Information
  Theory}, vol.~53, no.~1, pp. 394--401, 2006.

\bibitem{el2011network}
A.~El~Gamal and Y.-H. Kim, \emph{Network Information Theory}.\hskip 1em plus
  0.5em minus 0.4em\relax Cambridge, U.K.: Cambridge Univ. Presscfor, 2011.

\bibitem{han1981new}
T.~Han and K.~Kobayashi, ``A new achievable rate region for the interference
  channel,'' \emph{IEEE Transactions on Information Theory}, vol.~27, no.~1,
  pp. 49--60, 1981.

\bibitem{tandon2009outer}
R.~Tandon and S.~Ulukus, ``Outer bounds for multiple-access channels with
  feedback using dependence balance,'' \emph{IEEE Transactions on Information
  Theory}, vol.~55, no.~10, pp. 4494--4507, 2009.

\bibitem{el2005capacity}
A.~El~Gamal and S.~Zahedi, ``Capacity of a class of relay channels with
  orthogonal components,'' \emph{IEEE Transactions on Information Theory},
  vol.~51, no.~5, pp. 1815--1817, 2005.

\bibitem{kim2008capacity}
Y.-H. Kim, ``Capacity of a class of deterministic relay channels,'' \emph{IEEE
  Transactions on Information Theory}, vol.~54, no.~3, pp. 1328--1329, 2008.

\bibitem{chen2023general}
Y.~Chen, T.~Oechtering, M.~Skoglund, and Y.~Luo, ``On general
  capacity-distortion formulas of integrated sensing and communication,''
  \emph{arXiv preprint arXiv:2310.11080}, 2023.

\bibitem{nikbakht2024memory}
H.~Nikbakht, M.~Wigger, S.~S. Shitz, and H.~V. Poor, ``A memory-based
  reinforcement learning approach to integrated sensing and communication,'' in
  \emph{Proc. 58th Asilomar Conf. Signals, Syst, and Comput.}, 2024, pp.
  433--437.

\bibitem{lindstrom2025rate}
C.~P. Lindstrom and M.~R. Bloch, ``Rate distortion approach to joint
  communication and sensing with markov states: Open loop case,'' in
  \emph{Proc. IEEE ISIT}, 2025, pp. 1--6.

\bibitem{belghazi2018mutual}
M.~I. Belghazi, A.~Baratin, S.~Rajeshwar, S.~Ozair, Y.~Bengio, A.~Courville,
  and D.~Hjelm, ``Mutual information neural estimation,'' in \emph{Proc. ICML},
  2018, pp. 531--540.

\bibitem{gattegno2016fourier}
I.~B. Gattegno, Z.~Goldfeld, and H.~H. Permuter, ``Fourier-motzkin elimination
  software for information theoretic inequalities,'' \emph{arXiv preprint
  arXiv:1610.03990}, 2016.

\end{thebibliography}

\end{document}